\documentstyle[12pt,psfig]{article}
\input{epsf}
\newlength{\dinwidth}
\newlength{\dinmargin}
\begin{document}
 \newcommand{\ra}{\rightarrow}
 \newcommand{\as}{\mbox{$\alpha_{\displaystyle  s}$}}
 \newcommand{\mxtwo}{\mbox{$M^2_x$}}
 \newcommand{\Qtwo}{\mbox{$Q^2$}}
 \newcommand{\thetah}{\mbox{$\theta_H$}}
 \newcommand{\etamax}{\mbox{$\eta_{\rm max}$}}
 \newcommand{\betan}{\mbox{$\beta\;$}}
 \newcommand{\xbj}{\mbox{$\rm x_{Bj}$}}
\newcommand {\pom}  {I\hspace{-0.2em}P}
\newcommand {\xpom} {\mbox{$x_{_{\pom}}$}}
\newcommand {\xpomp}[1] {\mbox{$x^{#1}_{_{\pom}}\;$}}
\newcommand {\xpoma} {\mbox{$(1/x_{_{\pom}})^a\;$}}
\vspace{1 cm}
\title{
{\bf Measurement of the diffractive
     structure function in deep
     inelastic scattering at HERA}   \\
\author{\rm ZEUS Collaboration}
}
\date{ }
\maketitle
%---------------
\def\sleq {\raisebox{-.6ex}{${\textstyle\stackrel{<}{\sim}}$}}
\def\sgeq {\raisebox{-.6ex}{${\textstyle\stackrel{>}{\sim}}$}}
\def\F2Diff{\mbox{$F_2^{D(3)}$}}
\def\reg{{\cal R}}
\def\sm0{{ o}}
\def\mev{{\ \rm MeV}}
\def\gev{{\rm\  GeV}}
\def\fg#1{\noindent { figure #1}}
\def\f2g{$F_2^{\gamma}$}

\vspace{5 cm}
\begin{abstract}
\par\noindent
This paper presents an analysis of the inclusive properties of diffractive
deep inelastic scattering events produced in $ep$ interactions at HERA.
The events are characterised by a rapidity gap between the outgoing proton
system and the remaining hadronic system.  Inclusive distributions are
presented and compared with Monte Carlo models for diffractive processes.
The data are consistent with models where the pomeron structure function
has a hard and a soft contribution. The diffractive structure function is
measured as a function of $\xpom$, the momentum fraction lost by the
proton, of $\beta$, the momentum fraction of the struck quark with respect
to $\xpom$, and of $Q^2$. The $\xpom$ dependence is consistent with the
form \xpoma where $a~=~1.30~\pm~0.08~(stat)~^{+~0.08}_{-~0.14}~(sys)$ in
all bins of $\beta$ and $Q^2$.  In the measured $Q^2$ range, the
diffractive structure function approximately scales with $Q^2$ at fixed
$\beta$. In an Ingelman-Schlein type model, where commonly used pomeron
flux factor normalisations are assumed, it is found that the quarks within
the pomeron do not saturate the momentum sum rule.
\end{abstract}

\vspace{-20cm}
\begin{flushleft}
\tt DESY 95-093 \\
May 1995 \\
\end{flushleft}

\setcounter{page}{0}
\thispagestyle{empty}   % to suppress the page number on the first page
\newpage

\def\3{\ss}
\parindent 0cm
\footnotesize
\renewcommand{\thepage}{\Roman{page}}
\begin{center}
\begin{large}
The ZEUS Collaboration
\end{large}
\end{center}
M.~Derrick, D.~Krakauer, S.~Magill, D.~Mikunas, B.~Musgrave,
J.~Repond, R.~Stanek, R.L.~Talaga, H.~Zhang \\
{\it Argonne National Laboratory, Argonne, IL, USA}~$^{p}$\\[6pt]
R.~Ayad$^1$, G.~Bari, M.~Basile,
L.~Bellagamba, D.~Boscherini, A.~Bruni, G.~Bruni, P.~Bruni, G.~Cara
Romeo, G.~Castellini$^{2}$, M.~Chiarini,
L.~Cifarelli$^{3}$, F.~Cindolo, A.~Contin, M.~Corradi,
I.~Gialas$^{4}$,
P.~Giusti, G.~Iacobucci, G.~Laurenti, G.~Levi, A.~Margotti,
T.~Massam, R.~Nania, C.~Nemoz, \\
F.~Palmonari, A.~Polini, G.~Sartorelli, R.~Timellini, Y.~Zamora
Garcia$^{1}$,
A.~Zichichi \\
{\it University and INFN Bologna, Bologna, Italy}~$^{f}$
% ...
\\[6pt]
\clearpage
\hspace*{1mm}
$^{ 1}$ supported by Worldlab, Lausanne, Switzerland \\
\hspace*{1mm}
$^{ 2}$ also at IROE Florence, Italy  \\
\hspace*{1mm}
$^{ 3}$ now at Univ. of Salerno and INFN Napoli, Italy  \\
\hspace*{1mm}
$^{ 4}$ supported by EU HCM contract ERB-CHRX-CT93-0376 \\
\hspace*{1mm}
% ...
$^{18}$ now with OPAL Collaboration, Faculty of Physics at Univ. of
        Freiburg \\
% ...

\begin{tabular}{lp{15cm}}
$^{a}$ &supported by the Natural Sciences and Engineering Research
         Council of Canada (NSERC) \\
$^{b}$ &supported by the FCAR of Qu\'ebec, Canada\\
$^{c}$ &supported by the German Federal Ministry for Research and
         Technology (BMFT)\\
$^{d}$ &supported by the MINERVA Gesellschaft f\"ur Forschung GmbH,
         and by the Israel Academy of Science \\
$^{e}$ &supported by the German Israeli Foundation, and
         by the Israel Academy of Science \\
$^{f}$ &supported by the Italian National Institute for Nuclear Physics
         (INFN) \\
$^{g}$ &supported by the Japanese Ministry of Education, Science and
         Culture (the Monbusho)
         and its grants for Scientific Research\\
$^{h}$ &supported by the Korean Ministry of Education and Korea Science
         and Engineering Foundation \\
$^{i}$ &supported by the Netherlands Foundation for Research on Matter
         (FOM)\\
$^{j}$ &supported by the Polish State Committee for Scientific Research
         (grant No. SPB/P3/202/93) and the Foundation for Polish-
         German Collaboration (proj. No. 506/92) \\
$^{k}$ &supported by the Polish State Committee for Scientific
         Research (grant No. PB 861/2/91 and No. 2 2372 9102,
         grant No. PB 2 2376 9102 and No. PB 2 0092 9101) \\
$^{l}$ &partially supported by the German Federal Ministry for
         Research and Technology (BMFT) \\
$^{m}$ &supported by the German Federal Ministry for Research and
         Technology (BMFT), the Volkswagen Foundation, and the Deutsche
         Forschungsgemeinschaft \\
$^{n}$ &supported by the Spanish Ministry of Education and Science
         through funds provided by CICYT \\
$^{o}$ &supported by the Particle Physics and Astronomy Research
        Council \\
$^{p}$ &supported by the US Department of Energy \\
$^{q}$ &supported by the US National Science Foundation
\end{tabular}

\newpage
\pagenumbering{arabic}
\setcounter{page}{1}
\normalsize

\section{\bf Introduction}

We present an analysis of deep inelastic scattering (DIS) events with a
large rapidity gap between the outgoing proton system and the remaining
hadronic final state. The general properties of these events indicate that
the underlying production mechanism is leading twist and
diffractive~\cite{zeus1,zeus4}. Diffractive processes are generally
understood to proceed through the exchange of a colourless object with the
quantum numbers of the vacuum, generically called the pomeron~\cite{goul}.
The true nature of this exchanged ``object'' remains unclear.

The analysis of soft hadron-hadron collisions implies that
pomeron-exchange can be described by a pomeron-hadron coupling constant
and a pomeron-propagator \cite{lownus}.  This led to the proposition of
Ingelman and Schlein \cite{ingsch} to treat the pomeron as a quasi-real
particle which is emitted by a hadron, described in terms of a parton
density and characterised by a structure function $F_2^{\pom}$ which can
be studied in deep inelastic scattering.  The assumption of factorisation
implies that the pomeron structure is independent of the process of
emission.  Evidence for a partonic structure of the pomeron was observed
by the UA8 collaboration \cite{ua8} and later by the HERA experiments
\cite{zeus4,zeushpjet,h11,h1new}. These data also gave a first insight
into the structure function of the pomeron.  The UA8 data show a
predominantly hard structure, where on average the partons carry a large
fraction of the momentum of the pomeron. However, these data could not
distinguish between the quark and the gluon content of the pomeron.

This paper presents a study of the structure of the pomeron in DIS at
HERA. We discuss first the variables and cuts which are used to isolate
diffractive DIS events. The observed kinematic distributions of the data
are compared to Monte Carlo models of diffractive processes in a region
where the diffractive contribution dominates. A measurement of the
diffractive structure function is presented, integrated over $t$, the
square of the momentum transfer at the proton vertex, as a function of
$\xpom$, the momentum fraction lost by the proton, of $\beta$, the
momentum fraction of the struck quark with respect to $\xpom$, and of
$Q^2$. The data are used to determine the $\xpom$ dependence at fixed
$\beta$ in order to test factorisation; extract the $\beta$ dependence of
the diffractive structure function at fixed $Q^2$;  investigate the $Q^2$
dependence of the diffractive structure function at fixed $\beta$, in
order to test scale invariance; and, examine the general dependence on
$\xpom$, $\beta$ and $Q^2$ by comparing the data with different models for
diffractive dissociation.

\section{Experimental setup}

\subsection{HERA}

This analysis is based on data collected with the ZEUS detector at the
electron-proton collider HERA. During 1993, HERA was operated at a proton
energy, $E_p$, of 820 GeV and an electron energy, $E_e$, of 26.7 GeV. HERA
is designed to run with 210 bunches in each of the electron and proton
rings, with an interbunch spacing of 96 ns. For the 1993 data-taking 84
paired bunches were filled for each beam and in addition 10 electron and 6
proton bunches were left unpaired for background studies. Typical total
currents were 10 mA for both beams.

\subsection{The ZEUS detector}

Details of the ZEUS detector can be found in \cite{zeus2,zeus3}.  The
following is hence restricted to a short description of the components
relevant to the present analysis.

Charged particles are tracked by the inner tracking detectors which
operate in a magnetic field of 1.43 T provided by a thin superconducting
coil. Immediately surrounding the beampipe is the vertex detector (VXD)
which consists of 120 radial cells, each with 12 sense wires~\cite{vxd}.
Surrounding the VXD is the cylindrical central tracking detector (CTD)
which consists of 72 cylindrical drift chamber layers, organised into 9
superlayers~\cite{ctd} (5 axial and 4 small angle stereo layers). In
events with charged tracks, using the combined data from both chambers,
resolutions of $0.4$ cm in $Z$ and $0.1$ cm in radius in the $XY$
plane\footnote{The ZEUS coordinate system is defined as right handed with
the $Z$ axis pointing in the proton beam direction, hereafter referred to
as forward, the $X$ axis pointing towards the centre of HERA and the $Y$
axis pointing upwards.} are obtained for the primary vertex
reconstruction.

The high resolution uranium-scintillator
calorimeter (CAL)~\cite{test1,test2,pipe} consists of three parts,
forward (FCAL) covering the pseudorapidity\footnote{The pseudorapidity
$\eta$ is defined as $-\ln(\tan \frac{\theta}2)$, where the polar
angle $\theta$ is taken with respect to the proton beam direction and,
in this case, refers to the nominal interaction point.}
 region $4.3\geq \eta \geq 1.1$, barrel (BCAL) covering the central region
$1.1 \geq \eta \geq -0.75$ and rear (RCAL) covering the backward region
$-0.75\geq \eta \geq -3.8$. Holes of $20\times 20$ cm$^2$ in the centre of
FCAL and RCAL are required to accommodate the HERA beam pipe. The
resulting solid angle coverage is $99.7\%$ of $4\pi$.  The calorimeter is
subdivided longitudinally into one electromagnetic (EMC) section and one
(RCAL) or two (FCAL and BCAL) hadronic (HAC) sections.  The sections are
subdivided into cells, each of which is read out by two photomultiplier
tubes. The CAL also provides a time resolution of better than 1~ns for
energy deposits greater than 4.5~GeV, which is used for background
rejection.

The C5 beam monitor, a small lead-scintillator counter assembly around the
beam pipe located at $Z=-3.2$~m, has been used to measure the timing and
longitudinal structure of the proton and electron bunches, and to reject
events from upstream proton-gas interactions. The vetowall detector,
consisting of two layers of orthogonal scintillator strips on either side
of an 87 cm thick iron wall centred at $Z=-7.3$~m, was also used to tag
upstream background events.

The luminosity is measured from the rate observed in the luminosity photon
detector of hard bremsstrahlung photons from the Bethe-Heitler process $ep
\to e^\prime p\gamma$.  The luminosity detector \cite{lumi} consists of a
photon and an electron lead-scintillator calorimeter. Bremsstrahlung
photons emerging from the electron-proton interaction point at angles
below 0.5~mrad with respect to the electron beam axis hit the photon
calorimeter placed at 107~m distance along the electron beam line.
Electrons emitted at scattering angles less than 5~mrad and with energies
$0.2 E_e <E_e^\prime < 0.9 E_e$ are deflected by beam magnets and hit the
electron calorimeter placed 35~m from the interaction point.

\subsection{Trigger conditions}

Data were collected with a three level trigger \cite{zeus2}; details of
the first (FLT) and second (SLT) level decision for DIS events can be
found in previous publications \cite{F293}.

The third level trigger (TLT) rejects beam-gas events using timing and
cosmic-ray events by a combination of timing and topology.  Finally it
passes the accepted events through a set of filters in order to categorise
the events. The category of DIS neutral current events is defined by the
requirement of an electron candidate in the RCAL or BCAL. A cut was
performed on $\delta\equiv\Sigma_i
E_i(1-\cos\theta_i)>20\mbox{GeV}-2E_{\gamma}, $ where $E_i$ and $\theta_i$
are the energies and polar angles (in this case, with respect to the
nominal interaction point) of calorimeter cells and $E_{\gamma}$ is the
energy measured in the photon calorimeter of the luminosity monitor. For
events fully contained in the main detector $\delta\simeq 2E_e=53.4$ GeV,
whereas for low-$Q^2$ events the scattered electron escapes through the
rear beam pipe and $\delta$ peaks at low values.

\section{Kinematic variables}
\label{s:kin}

The kinematic variables used to describe DIS events $$e~(k)~+~ p~(P)
\rightarrow e'~(k')~+~anything $$ are the following:  the negative of the
squared four-momentum transfer carried by the virtual photon\footnote{In
the $Q^2$ range used for this analysis, $ep$ interactions are described to
sufficient accuracy by the exchange of a virtual photon.}:  $$
Q^2=-q^2=-(k~-~k')^2; $$ the Bjorken variable:  $$ x =\frac{Q^2}{2P\cdot
q}; $$ the variable which describes the energy transfer to the hadronic
final state:  $$ y =\frac{P\cdot q}{P\cdot k}; $$ and the centre-of-mass
energy $W$ of the virtual-photon proton ($\gamma^*p$) system, where:  $$
W^2=(q+P)^2=\frac{Q^2(1-x)}{x}+M_p^2 $$ with $M_p$ denoting the proton
mass.

These variables, only two of which are independent at fixed $ep$
centre-of-mass energy squared $s=(k+P)^2$, can be reconstructed in a
variety of ways using combinations of electron and hadronic system
energies and angles~\cite{bent}. The variable $y$, calculated from the
electron variables, is given by: $$ y_e = 1 -
\frac{E^{\prime}_e}{E_e}~\frac{1-\cos \theta^{\prime}_e}{2} $$ where
$E^{\prime}_e$, $\theta^{\prime}_e$ denote the energy and angle of the
scattered electron.  Alternatively, $y$ can be estimated from the hadronic
system, using the Jacquet-Blondel technique \cite{jblo}:  $$
y_{JB}=\frac{\sum_i E_i(1 - cos\theta_i)}{2\cdot E_e} $$ where $E_i$ and
$\theta_i$ are the energies and polar angles of calorimeter cells which
are associated with the hadronic system.

Studies of the kinematic variables have shown that for this analysis it is
advantageous to use the so-called double angle ($DA$) method, in which the
angles of the scattered electron and the hadron system are used to
determine $x$ and $Q^2$.  Quantities determined in this way will be
denoted by the subscript $DA$. Formulae to calculate $Q^2_{DA}$, $W_{DA}$,
$x_{DA}$ and $y_{DA}$ are given in~\cite{bent}.

In the diffractive DIS process shown in Fig.~1: $$ e~(k)~+~ p~(P)
\rightarrow e'~(k')~+~p'(P')~+~X,$$ the hadronic system $X$ (exclusive of
the proton) and the scattered electron $e'$ are detected in the main
detector. The proton remnant $p'$ remains undetected. When the system $X$
is fully contained its invariant mass, $M_X$, can be determined from the
calorimeter cell information as follows~\cite{zeus1}.  Denoting the
energy, momentum and polar angle of the final hadronic system as $E_H$,
$p_H$ and $\theta_H$, respectively; and $\vec{p_i}$ as the vector
constructed from the energy $E_i$, polar angle $\theta_i$ and azimuthal
angle $\phi_i$ of cell $i$; then: \begin{equation}
\cos\theta_H=\frac{\sum_i p_{z_i}}{|\sum_i \vec{p_i}|} \label{eq:hadangle}
\end{equation} $$ p_H^2={Q^2_{DA}(1-y_{DA})\over \sin^2\theta_H} $$ $$
E_H=p_H\cos\theta_H+2E_ey_{DA} $$ from which $M_X$ is determined by the
definition $M_X=\sqrt {E_H^2 - p_H^2}$.

The squared four-momentum transfer at the proton vertex is given by: $$ t
= (P-P')^2,$$ whose absolute magnitude is expected to be small compared to
$Q^2+M_X^2$ in diffractive processes for the kinematic region studied
here. To describe diffractive deep inelastic scattering, in addition to
$x$ and $Q^2$, the following variables are used: $$ \xpom =
\frac{(P-P')\cdot q}{P\cdot q} = \frac{M_X^2 + Q^2 - t} {W^2 + Q^2 -
M_p^2} \simeq \frac{M_X^2 + Q^2}{W^2 + Q^2},$$ $$ \beta =
\frac{Q^2}{2(P-P')\cdot q} = \frac{x}{\xpom} = \frac{Q^2}{M_X^2 + Q^2 - t}
\simeq \frac{Q^2}{M_X^2 + Q^2}.$$ In models where diffraction is described
by the exchange of a particle-like pomeron, $\xpom$ is the momentum
fraction of the pomeron in the proton and $\beta$ is the momentum fraction
of the struck quark within the pomeron. For the structure of the pomeron
in DIS, the variable $\beta$ plays a role analogous to that of Bjorken-$x$
for the structure of the proton.

\section{Diffractive structure function}
\label{s:dsf}

For unpolarised beams, the differential cross section for single
diffractive dissociation can be described in terms of the diffractive
structure function, $F_2^{D(4)}(\beta,Q^2,\xpom,t)$: $$
\frac{d^4\sigma_{diff}}{d\beta dQ^2 d\xpom dt} = \frac{2\pi
\alpha^{2}}{\beta Q^{4}} \; [(1+(1-y)^2) F_{2}^{D(4)} - y^2
F_{L}^{D(4)}]\;  (1+\delta_{Z}) (1 + \delta_{r}) $$ where $\alpha$ is the
electromagnetic coupling constant and the $\delta_i$ denote corrections
due to $Z^0$ exchange and due to radiative corrections which are small in
the measured range. The contribution of $F_L$ to the diffractive cross
section is not known. If such a term were included, the $F_2$ values would
become larger at large $y$ values (corresponding to small $\xpom$ values).
The effect of this uncertainty is considered in section~\ref{s:f2d}. Note
that the function $F_2^{D(4)}(\beta,Q^2,\xpom,t)$ can be related to that
of ${\cal F}_2^{D(4)}(x,Q^2,\xpom,t)$. Integrating ${\cal F}_2^{D(4)}$
over $\xpom$ and $t$ one can directly compare it to the inclusive proton
structure function $F_2(x,Q^2)$~\cite{buch2}.

In this analysis, an integral is performed over $t$, corresponding to the
(undetected) momentum transfer to the proton system. For this initial
measurement we neglect the effect of $F_L$ and the additional
contributions noted above, yielding the following expression for \F2Diff,
where the cross section is evaluated as a function of $\beta,Q^2$ and
$\xpom$:
$$
  \frac{d^3\sigma_{diff}}{d\beta dQ^2 d\xpom} = \frac{2 \pi
    \alpha^2}{\beta Q^4} \; (1+(1-y)^2) \;
  F_2^{D(3)}(\beta,Q^2,\xpom),
$$
following the procedure of~\cite{ingprytz}, where the relation
$x = \beta \xpom$ has been used.

\section{Diffractive models and Monte Carlo simulation}
\label{s:MC}

\begin{figure}[ht]
\begin{center}
\leavevmode
\hbox{%
\epsfxsize = 4.5in
\epsffile{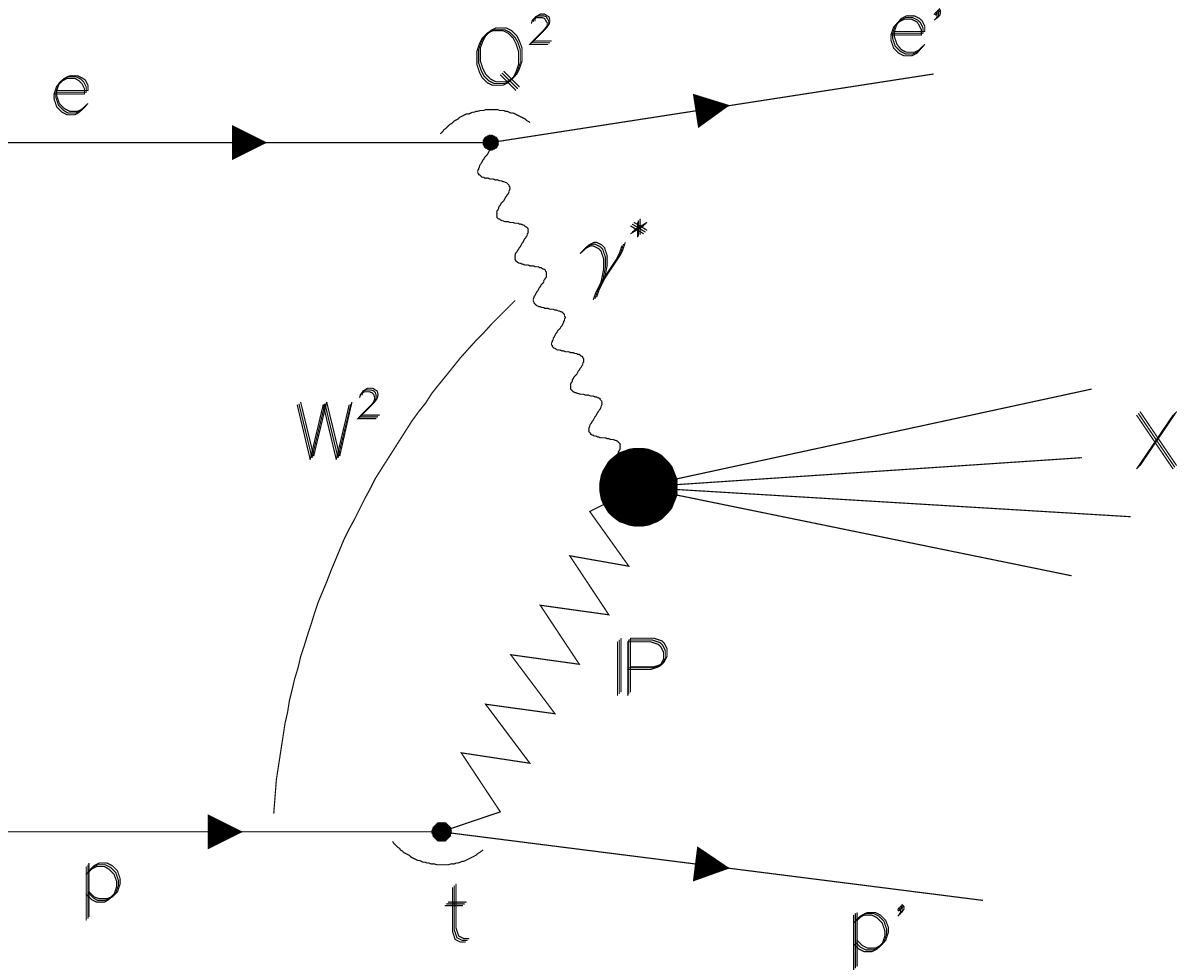}}
\end{center}
\caption{{\protect\small Diagram of a diffractive event.}}
\label{diffr}
\end{figure}

Different approaches exist to model diffractive processes such as that
depicted in Fig.~\ref{diffr}. In this paper we compare the data with the
predictions of the factorisable models of Ingelman and
Schlein~\cite{ingsch}, Donnachie and Landshoff~\cite{dl1} and Capella et
al.~\cite{capella}, as well as the non-factorisable model of Nikolaev and
Zakharov~\cite{nz,nznew}. We have earlier found that the Monte Carlo
implementations~\cite{pompyt,ada} of the models described above provide
reasonable descriptions of the shape of the energy flow and of the
observed fraction of events with one or more jets~\cite{zeus4,zeus5}.

In the model of Ingelman and Schlein \cite{ingsch} the proton emits a
pomeron which is treated as a (virtual) hadron whose structure is probed
by the virtual photon. The pomeron is described by a structure function
$F_2^{\pom}(\beta,Q^2)$ which is independent of the process of emission.
In this sense factorisation is predicted in the model: $$
F_2^{D(4)}(\beta, Q^2, \xpom, t)
   = f_{\pom}(\xpom,t) \cdot F_2^{\pom}(\beta,Q^2). $$

The flux factor $f(\xpom,t)$, describing the flux of pomerons in the
proton, can be extracted from hadron-hadron scattering with an accuracy of
approximately 30\%, assuming universality of the pomeron flux.  A
comparison of different flux factors can be found in \cite{kjell}.

For this analysis we used the POMPYT Monte Carlo implementation
\cite{pompyt} of the Ingelman-Schlein model. Two samples of events were
generated, corresponding to a hard quarkonic structure function, $$
F_2^{\pom}(\beta, Q^2) = \sum_{q_i} e_i^2 \beta f_q(\beta, Q^2) =
\frac{5}{3} \cdot \beta(1-\beta), $$ and to a soft quarkonic structure
function, $$ F_2^{\pom}(\beta, Q^2) = \sum_{q_i} e_i^2 \beta f_q(\beta,
Q^2) = \frac{5}{3} \cdot (1-\beta)^5. $$ The two samples are denoted by
``Hard Pomeron'' (HP) and ``Soft Pomeron'' (SP) respectively. The
normalisation constant 5/3 is based on the assumption that the momentum
sum rule (MSR) is satisfied for two light quark flavours (u,d).  If s
quarks would have to be included the normalisation factor would be reduced
from 5/3 to 4/3~\cite{stirling}. The $Q^2$ dependence is expected to be
weak and is neglected.  The Ingelman-Schlein form of the flux is
parametrised by a fit to UA4 data~\cite{pompyt,ua4}: $$ f_{\pom}(\xpom,t)=
\frac{1}{2} \frac{1}{2.3 \cdot \xpom} \cdot (6.38~e^{8t} +
0.424~e^{3t}).$$

In the Donnachie-Landshoff (DL) model diffraction in DIS is described
through pomeron exchange between the virtual photon and the proton, with
the pomeron coupling predominantly to quarks~\cite{dl2}.  The authors
calculate the cross section in the framework of Regge theory. The result
can be interpreted in terms of a pomeron structure function with the
resulting $\beta$ dependence similar to HP but with a normalisation which
is calculated to be approximately a factor of 6.2 smaller. The authors
also predict an additional soft contribution to the pomeron structure
function which is expected to become important only for $\beta < 0.1$. The
flux factor, $$ f_{\pom}(\xpom,t)=\frac{9\beta_0^2}{4\pi^2} F_1(t)^2
\xpom^{1-2\alpha(t)}, $$ is related to the elastic form factor of the
proton, $F_1(t) = \frac{4M^2 - 2.8t}{4M^2-t}(\frac{1}{1 - t/0.7})^2$, and
to the pomeron-quark coupling, $\beta_0 \simeq 1.8 ~{\rm GeV^{-1}}$,
extracted from hadron-hadron data.  The $\xpom$ term represents the
pomeron propagator with the pomeron trajectory, \mbox{$\alpha(t) = 1.085 +
0.25 \cdot t$.} Therefore the $\xpom$ dependence of $F_2^{D(4)}$ is
controlled by the pomeron trajectory
($F_2^{D(4)}~\propto~\xpom^{1-2\alpha(t)}$).  Integrated over $t$, the
predicted effective $\xpom$-dependence at fixed $\beta$ is approximately
$(1/\xpom)^{1.09}$ in the measured range of $\xpom$.

In a recent publication~\cite{goulnew} Goulianos proposed to use a
modified flux factor, which is renormalised to unity for fixed
centre-of-mass energy $W$. Also a modification of the $t$-dependence and
the pomeron trajectory according to recent CDF data~\cite{CDF} is
proposed: $$ f_{\pom}(\xpom,t)= \frac{1}{N_0} \cdot 0.73 \cdot e^{4.6t}
\cdot \xpom^{1-2\alpha(t)}, $$ where $\alpha(t) = 1.115 + 0.26 \cdot t$
and $N_0$ is a normalisation factor which can be approximated by
$(\frac{W^2}{400})^{0.23}$.  Integrated over $t$, the effective
$\xpom$-dependence of this flux factor is approximately
$(1/\xpom)^{0.93}$.

Capella et al. calculate the diffractive structure function in the
framework of conventional Regge theory \cite{capella}.  Using Regge
factorisation, they relate the pomeron structure function to the deuteron
structure function using parameters which are determined from soft
hadronic diffraction data with an appropriate change for the disappearance
of screening corrections with increasing $Q^2$. For $Q^2 = 10$~GeV$^2$
they obtain: $$ F_2^{\pom}(\beta, Q^2) = a \cdot \beta^{0.6} \cdot
(1-\beta)^{0.6} + 0.015 \cdot \beta^{-0.22} \cdot (1-\beta)^{4.6} \ ,$$
where $a$ is estimated to be in the range 0.04 to 0.06. We chose $a =
0.06$ for comparison with the data.

In the model of Nikolaev and Zakharov diffractive dissociation is
described as a fluctuation of the photon into a $q\bar{q}$ or $q\bar{q}g$
Fock state~\cite{nz,nznew}.  The interaction with the proton proceeds via
the exchange of a BFKL~\cite{LIP} type pomeron, starting in lowest-order
from the exchange of a Low-Nussinov~\cite{lownus} pomeron which
corresponds to two gluons in a colour-singlet state. The result for the
cross section can be approximated by a two-component structure function of
the pomeron, each component having its own flux factor. This corresponds
to factorisation breaking which is caused by BFKL evolution effects.  The
result for the ``hard" component reflects the case where the photon
fluctuates into a $q\bar{q}$ pair and leads to a $\beta$ dependence
similar to that of the HP and the DL models with a normalisation closer to
the latter (but with very different predictions for the contribution of
heavy flavours). The ``soft" contribution, which reflects the case where
the photon fluctuates into $q\bar{q}g$, is assumed to be proportional to
$(1-\beta)^2$ and the normalisation is fixed by the triple pomeron
coupling.  The relative size of these contributions and the overall
normalisation are predicted with an uncertainty of about 30\%. $Q^2$
evolution effects have been calculated for this model, and have been found
to be rather weak for $\beta > 0.1$. We used a Monte Carlo implementation
of this model~\cite{ada} which is based on the cross section given in
\cite{nz} and is interfaced to the Lund fragmentation scheme. We refer to
this model as NZ. In this implementation the mass spectrum contains both
components but the $q\bar{q}g$ states are fragmented into hadrons as if
they were a $q\bar{q}$ system with the same $M_X$.

Both Monte Carlo generators have limitations in the generation of small
masses $M_X$ \linebreak ($M_X~<~1.7$~GeV for NZ and $M_{X^\prime} < 5$~GeV
for POMPYT, where $M_{X^\prime}$ includes the final state electron). We
exclude these regions for the measurement of the diffractive structure
function by an upper cut on $\beta$.  To study acceptance and migration
effects for these small masses we have generated an additional sample of
exclusively produced $\rho^0$'s~\cite{rhopaper}.

The cuts given below to select diffractive events limit the acceptance for
double-dissociative events, where the proton also dissociates.  The PYTHIA
Monte Carlo~\cite{PYTHIA} has been used to study the detector response for
the nucleon system $M_N$ in double dissociation $\gamma p$ events. The
nucleon system mass spectrum and the fraction of double-dissociative to
single-dissociative events was taken from hadron data.

Non-diffractive DIS processes were generated using the HERACLES~4.4
\mbox{program}~\cite{HERACLES} which incorporates first order electroweak
corrections.  The Monte Carlo generator LEPTO~6.1 \cite{LEPTO}, interfaced
to HERACLES via the program DJANGO~6.0~\cite{DJANGO}, was used to simulate
QCD cascades and fragmentation.  The parton cascade was modelled with the
colour-dipole model including the boson-gluon fusion process by the
ARIADNE~4.03~\cite{ariadne} (CDMBGF) program.  The fragmentation into
hadrons was performed with the Lund string hadronisation model \cite{lund}
as implemented in JETSET~7.2~\cite{jetse}.  For the proton parton
densities the $\rm{MRSD_-^{\prime}}$ set \cite{martin} was chosen, which
adequately represents our structure function results \cite{F293}.  In
previous studies \cite{zeus5} it was shown that this model gives a good
description of the energy flow between the current jet and the remnant
jet.

Combined sets of the Monte Carlo generators, described above, were used to
simulate the expected final states in our DIS sample:  the first one to
describe non-diffractive DIS processes and the second one to model
diffractive events.

QED radiative processes were not simulated for diffractive events;
however, with the selection cuts of section~\ref{s:selec}, radiative
corrections to the DIS cross sections are below 10\% \cite{F293} and are
expected to be of the same order for diffractive processes.  All Monte
Carlo events were passed through the standard ZEUS detector and trigger
simulations and the event reconstruction package \cite{zeus2}.  According
to Monte Carlo studies, the efficiency of the trigger and of the final
selection cuts is the same for diffractive and non-diffractive DIS events.
The overall trigger acceptance is above 95\%, independent of $x$ and $Q^2$
in the range of interest for this analysis.

\section{Event selection}
\label{s:selec}

The selection of DIS events was similar to that described in our
earlier publications \cite{zeus1,F293}.
The following offline cuts were applied:
\begin{itemize}
\item E$_{e}^\prime \geq$ 5 GeV, to ensure good electron identification;
\item $Q^2_{DA}\geq 8$ GeV$^2$;
\item $y_{JB}\geq 0.04$,
to give sufficient accuracy for DA reconstruction;
\item $\delta \geq 35$ GeV (with respect to the measured interaction
point), to reduce radiative corrections and photoproduction background;
\item $y_e\leq 0.95$, to reduce photoproduction background;
\item the impact point of the electron on the face of the RCAL was required
to lie outside a square of side 32 cm centred on the beam axis (box cut),
to ensure that the electron shower was fully contained within the calorimater
and its position could be reconstructed with sufficient accuracy;
\item a vertex, as reconstructed from VXD+CTD tracks, was required with
$|Z_{vtx}|\leq 40$ cm.
\end{itemize}
In addition algorithms were used to reject cosmic-ray induced events
and QED Compton events.

A total of 31k events was selected in this way corresponding to an
integrated luminosity of 0.54~pb$^{-1}$.  Using the number of events
produced by unpaired electron and proton bunches, the contamination from
beam-gas background and from cosmic-ray muons were estimated to be less
than 1\% each. The background in the total DIS sample due to
photoproduction was estimated to be ($2.5\pm1$)\% from a fit to the shape
of the $\delta$ distribution before the above cut on $\delta$ was applied
\cite{F293}. %This results in 30306$\pm$ 174 (stat)$\pm$325 (sys) events.

\section{\bf Properties of diffractive events}

In the following section, the observed data distributions are compared to
distributions from various Monte Carlo models. We discuss the criteria
used to select the diffractive events, the methods used to determine their
relative contribution and the observed inclusive distributions of the
diffractive sample.

\subsection{\bf Selection criterion}

A large fraction of diffractive processes at HERA exhibit a rapidity gap
in the main detector between the scattered proton system and the hadronic
activity generated by the dissociation of the photon, while large rapidity
gaps are suppressed in non-diffractive DIS events.  Therefore the presence
of a rapidity gap has been used as a selection criterion
\cite{zeus1,zeus4}. An improved criterion to separate diffractive from
non-diffractive events is presented here, which uses the direction of the
total hadronic energy flow of the event, determined from all the detected
particles in the final state.

We define the maximum pseudorapidity of an event, ${\eta_{\rm max}}$, as
the maximum value of the pseudorapidity of all calorimeter condensates
with energy greater than 400 MeV or tracks with momentum of at least 400
MeV/c. A condensate is a contiguous energy deposit above a minimum energy
threshold. We studied the effect of varying the minimum energy $E_{\rm
min}$ = 400 MeV and found that above ${E_{\rm min}}=200$ MeV the ${\rm
\eta_{\rm max}}$ distribution does not change significantly.  We chose
$E_{\rm min}$ = 400~MeV as a conservative compromise between accepting
diffractive events and rejecting noise.

For values of \etamax\ up to 1--1.5 the non-diffractive DIS background is
a negligible background to the diffractive sample, which increases for
values of \etamax\ above 1.5--2. In previous ZEUS publications
\cite{zeus1,zeus4} diffractive events were selected by $\etamax < 1.5$.
This cut selects a rather pure sample of diffractive events, useful to
establish a signal but it limits acceptance for events with large $M_X$.

\begin{figure}[htb]
\begin{center}
\leavevmode
\hbox{%
\epsfxsize = 6.in
\epsffile{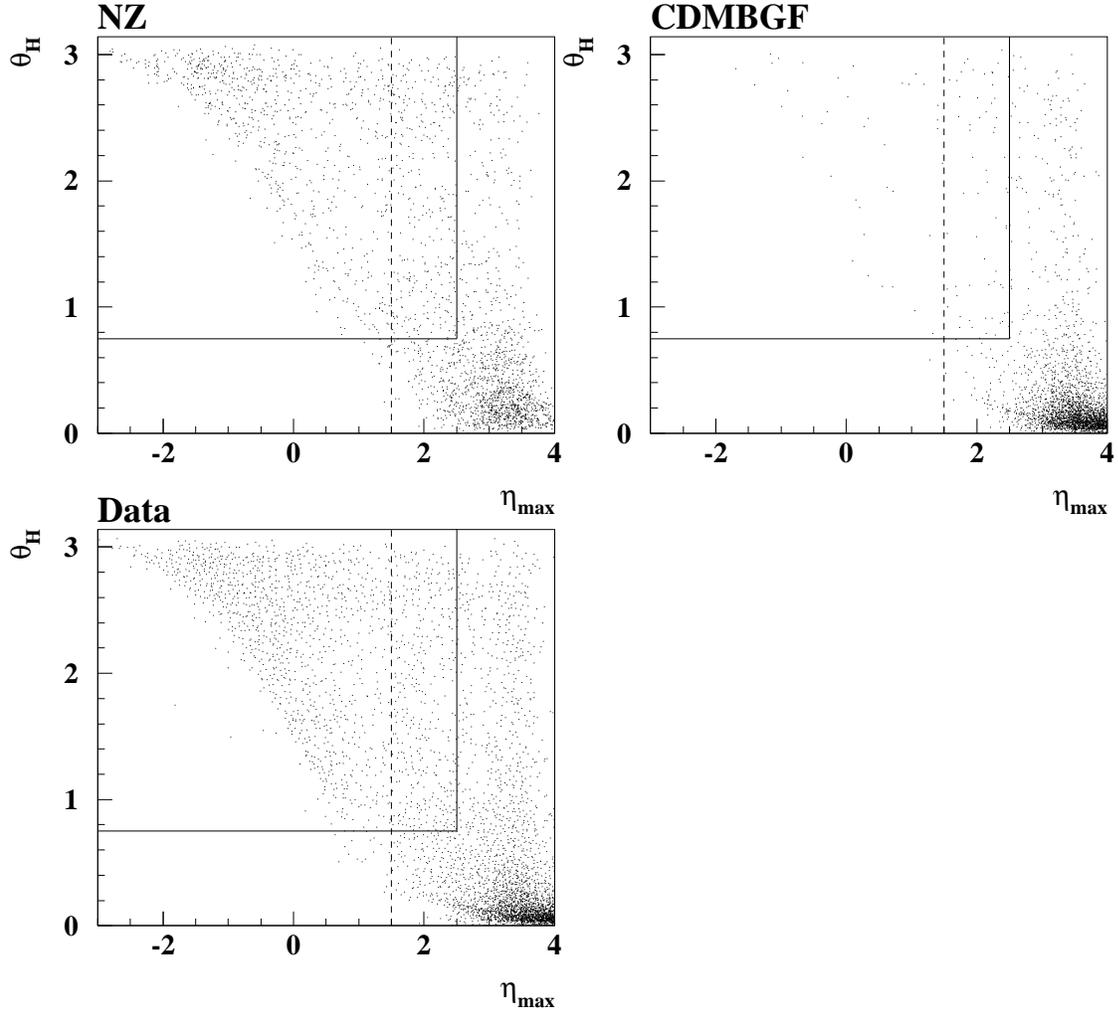}}
\end{center}
\caption{{\protect\small $\theta_{H}$ versus
${\eta_{\rm max}}$ distribution for diffractive (NZ) and
non-diffractive (CDMBGF) Monte Carlo events and for the selected DIS data.
The full line indicates the \etamax-\thetah~cut used to select diffractive
events. The dotted line corresponds to \etamax=1.5.
}}
\label{coseta}
\end{figure}

The \etamax-cut is dependent on the most forward condensate but does not
use the information from the full energy flow. Larger acceptance can be
achieved by including more information from the hadronic energy flow.
Since in diffractive scattering the proton remains intact or, in the case
of double-dissociative events, dissociates independently from the photon,
the hadronic activity in the detector in general will not follow the
proton direction. The hadronic angle \thetah\ defined in
eq.~(\ref{eq:hadangle}) represents the average direction of the hadronic
activity. Non-diffractive DIS events have mostly cos$\theta_H$ near~1
because of the colour flow between the struck quark and the outgoing
proton system, while a substantial fraction of diffractive events is found
at cos$\theta_H$ less than~1. Figures~\ref{coseta}a, b show scatter plots
of $\eta_{\rm max}$ versus $\theta_H$ for the diffractive and
non-diffractive DIS Monte Carlo samples. A cut cos$\theta_H<0.75$ combined
with ${\eta_{\rm max}<2.5}$ allows a larger acceptance of diffractive
events than the ${\eta_{\rm max}<1.5}$ cut, at the price of a slightly
higher background which has to be subtracted. We call this combined cut,
used to select the diffractive sample, the \etamax-\thetah~cut.

\begin{figure}[bht]
\begin{center}
\leavevmode
\hbox{%
\epsfxsize = 4.5in
\epsffile{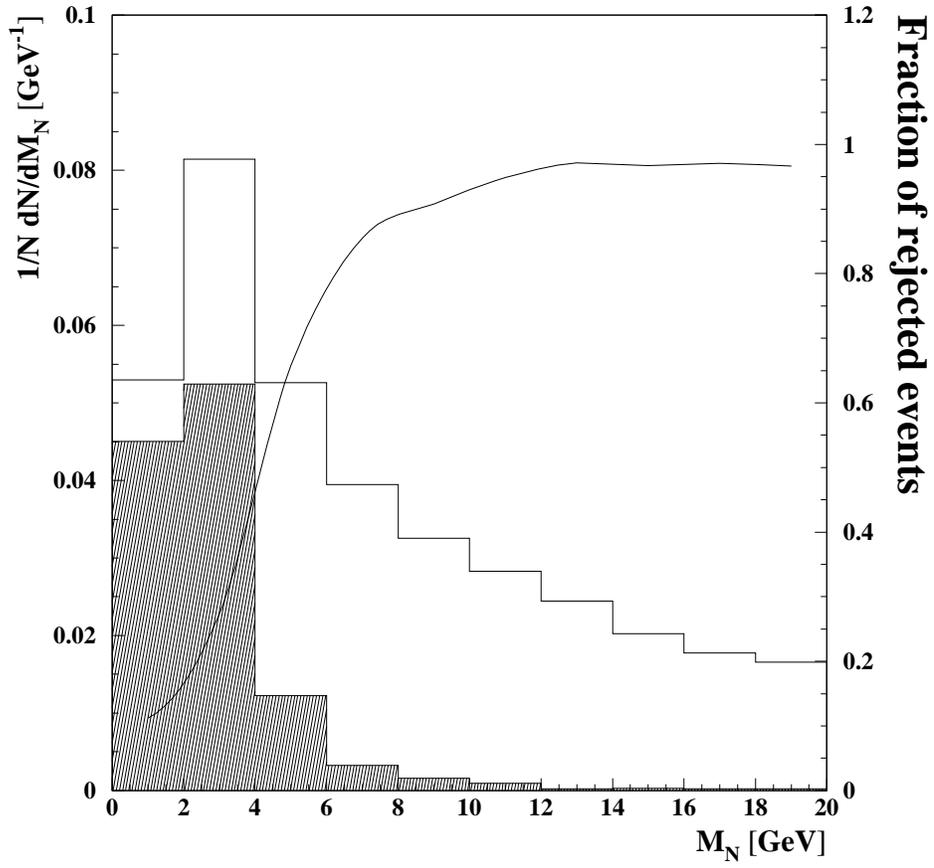}}
\end{center}
\caption{{\protect\small
Acceptance for double dissociative events.
The mass of the nucleon system, $M_{\rm N}$, for double dissociative events
generated by the PYTHIA Monte Carlo is indicated by the full line histogram.
The shaded area indicates those events which are selected by the
\etamax-\thetah~cut.
The fraction of double dissociative
events rejected by this cut, as a function of $M_{\rm N}$,
is indicated by the line.}}
\label{ddacc}
\end{figure}

Note that the term ``diffractive" is used to indicate single diffractive
dissociation of the photon together with that fraction of events where
both the photon and the proton dissociate and the proton system is not
detected. From proton-proton measurements of the ratio of double- to
single-dissociative events, we estimate this ratio to be approximately
0.76 in the measured $W$ range. As shown in Fig.~\ref{ddacc}, we find that
excited proton states with mass $M_N~\sleq~4$~GeV would pass the
diffractive selection cuts. Beyond this range the energy deposition in the
forward calorimeter is typically above 400 MeV. The overall acceptance for
double-dissociative events is 23\%. We therefore have an estimated
double-dissociative contribution of $\simeq$~(15$\pm$10)\% which is
expected to be independent of $\beta$ and $Q^2$ and not vary significantly
with $\xpom$. This result assumes factorisation in Regge theory, i.e. that
the nucleon mass spectrum and the ratio of double- to single-dissociative
events is similar to that measured in proton-proton collisions at similar
energies.

\subsection{Estimation of the diffractive component}

In this section, only the shapes of the distributions and not the absolute
normalisations of the diffractive models are considered. The ${\eta_{\rm
max}}$ and $\theta_H$ distributions are used to determine the fraction of
diffractive events passing the DIS selection criteria. A linear
combination of diffractive DIS (NZ or POMPYT) and non-diffractive DIS
Monte Carlo events are fitted to the data.

\begin{figure}[htb]
\begin{center}
\leavevmode
\hbox{%
\epsfxsize = 6.in
\epsffile{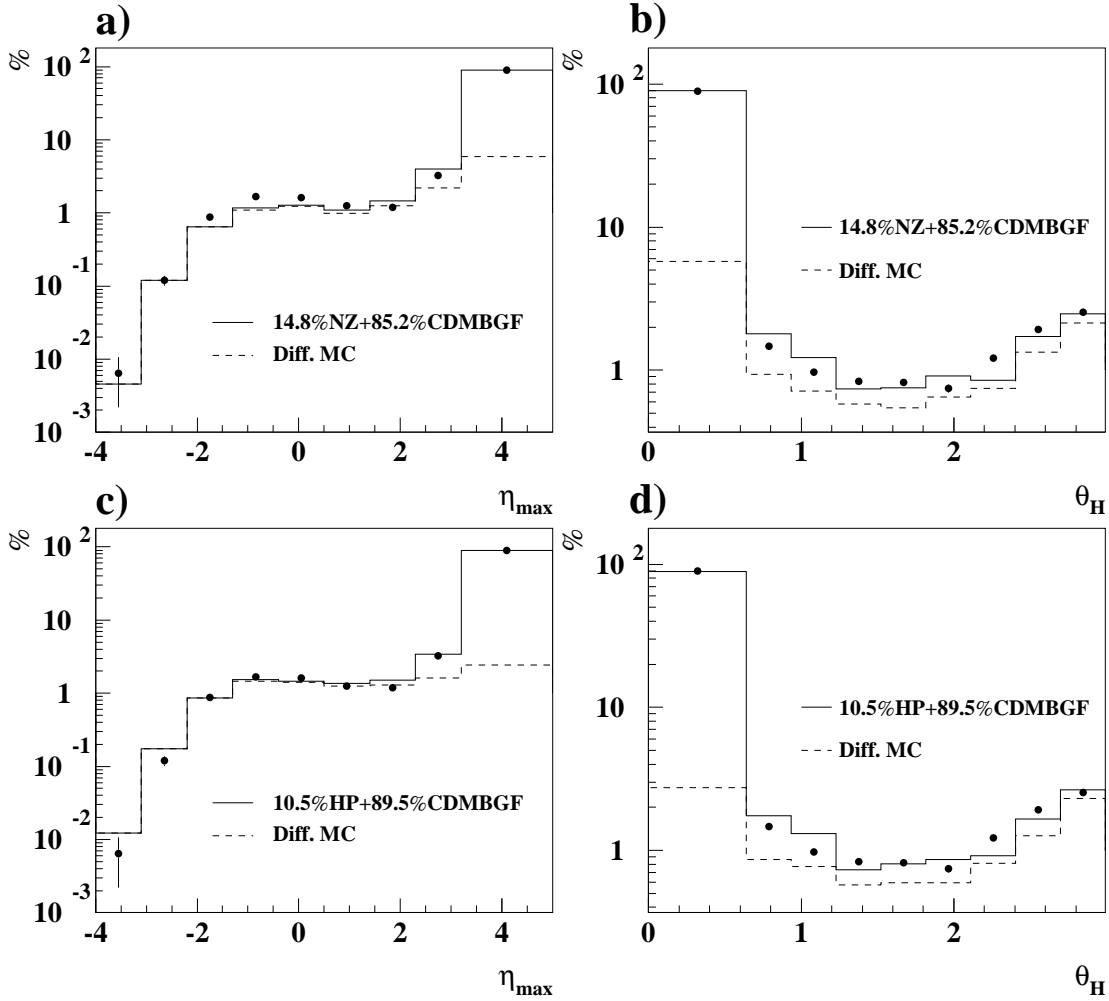}}
\end{center}
\caption{{\protect\small  Percentage of DIS data as a function
of \etamax\ and \thetah.
%The data are compared to the fitted estimate from Monte Carlos.
The data are described by the sum of the diffractive and non-diffractive
contributions obtained from Monte Carlo simulation, with relative
fractions determined by a fit to the data.
The dashed line corresponds to the diffractive contribution
and the sum of the diffractive and non-diffractive Monte Carlo models
is indicated by the full line.}}
\label{fit3}
\end{figure}

The ${\eta_{\rm max}}$ and $\theta_H$ distributions were first fitted
separately to check consistency between the results and then together to
obtain a global result.  Figure~\protect\ref{fit3} shows the fits to these
distributions. The part of each distribution that corresponds to the
forward region of the detector (high values of \etamax\ and low values of
\thetah) was put into one single bin to reduce problems associated with a
detailed description of the hadronisation of the proton remnant. For each
distribution a variety of different binnings was tried and the results
were found to be stable.

\begin{table}[htb]
\begin{center}
\begin{tabular}{|l|l|l|l|l||l|l|}  \hline
 Model  &\multicolumn{2}{c|}{\etamax}&\multicolumn{2}
{c||}{\thetah}&\multicolumn{2}{c|}{\etamax+\thetah} \\ \hline
 & \% of diffr.&$\chi^2_{dof}$ & \% of diffr.&$\chi^2_{dof}$& \% of diffr.&
$\chi^2_{dof}$ \\ \hline \hline
%without rho's
NZ& 14.2 $\pm$ 2.5 & 4.7  & 15.3 $\pm$ 2.5 &  4.0& 14.8 $\pm$ 3.0 & 4.3 \\
\hline
SP& 35.9 $\pm$ 7.0 & 10.5 & 15.4 $\pm$ 2.6 &  5.0& 33.0 $\pm$ 6.0  & 24 \\
\hline
HP& 10.3 $\pm$ 2.0 & 2.2  & 10.8 $\pm$ 2.0 & 4.1& 10.5 $\pm$ 2.7& 3.0 \\ \hline
HP+SP& 15.6 $\pm$ 1.3 & 3.7 & 13.4 $\pm$ 1.3 & 5.4& 14.6 $\pm$ 1.4& 4.6 \\
\hline
%with rho's
%NZ& 14.4 $\pm$ 2.5 & 2.0  & 15.3 $\pm$ 2.5 &  1.5& 14.8 $\pm$ 3.0 & 1.8 \\
%%\hline
%SP& 35.9 $\pm$ 7.0 & 10.5 & 15.4 $\pm$ 2.6 &  5.0&  - & - \\ \hline
%HP& 10.0 $\pm$ 2.0   & 1.1  & 11.0 $\pm$ 2.0 & 1.4& 10.5 $\pm$ 2.7& 1.2 \\
%%\hline
\end{tabular}
\caption{{\protect\small Fraction of diffractive events and
$\chi^2$ per degree of freedom ($\chi^2_{dof}$) values obtained from fits
using NZ, SP, HP or HP+SP.}}
\end{center}
\end{table}

The results are summarised in Table~1, where the default parameters have
been used for the models. Since neither NZ nor POMPYT describes
diffractive vector meson production a simulation of exclusive $\rho^o$
production was added in order to incorporate the effect of low-mass
states.  This contribution was estimated to be typically $\sim$~7\% of the
diffractive sample from a fit to the observed $M_X$ spectrum in different
$Q^2$ intervals. For each model, a reduction of $\chi^2_{dof}$ by~1--2 was
found when $\rho^o$ production was included, with consistent results
obtained for the fraction of diffractive events.

The SP model was also extensively tested. In fits to the \etamax\ and
\thetah\ distributions, SP does not reproduce the shapes correctly.  Its
very soft $\beta$ distribution tends to populate large \etamax\ bins and,
consequently, the fits do not describe the data. The inconsistency of the
results obtained by the fits to the \etamax\ and the \thetah\
distributions shown in Table~1 indicate that a pure soft \betan
distribution cannot describe the data. For these reasons the SP model
alone is not considered any further. Results obtained with the combined
HP+SP model, discussed in the following section, are also given in
Table~1. The fractions obtained with the HP+SP model are similar to those
determined using the NZ model.

It is possible to explain the different predictions from NZ and HP models
in terms of the \betan distribution used in these models: they both
contain the hard component responsible for low $\etamax$ (high $\thetah$)
events but the NZ model also contains a soft contribution which is
predicted to be $\sim$ 40\% of the diffractive cross section.  Most of the
events originating from this soft component are hidden under the large
background from normal DIS events and so are not accessible to the present
study.

\subsection{Inclusive distributions}

In the following, the shapes of the observed distributions in $W$, $Q^2$,
$x$, $M_X$, $\xpom$ and $\beta$ are considered.  The relative
normalisation of the models is obtained from the above fits. It should be
noted that the normalisation of the non-diffractive component, which is
relevant for the background subtraction, is independent of the diffractive
model used to fit the data to within 5\%.

In order to confine the analysis to regions of acceptance above $\simeq$~80\%,
the following ($M_X,y$) intervals were considered:
\begin{eqnarray*}
  M_X < 10~{\rm GeV}\ & {\rm for}\ & 0.08 < y < 0.2 \\
  M_X < 16~{\rm GeV}\ & {\rm for}\ & 0.2 < y < 0.3 \\
  M_X < 20~{\rm GeV}\ & {\rm for}\ & 0.3 < y < 0.8
\end{eqnarray*}
According to Monte Carlo
studies, the \etamax-\thetah~cut reduces the non-diffractive DIS
component by $\simeq$ 60\% and the diffractive
component by $\simeq$ 20\%, giving a contamination from non-diffractive DIS
of less than 15\% in these ($M_X$, $y$) intervals.
This background is subtracted from the data
before comparison with the diffractive Monte Carlo
predictions.

\begin{figure}[htb]
\begin{center}
\leavevmode
\hbox{%
\epsfxsize = 5.5in
\epsffile{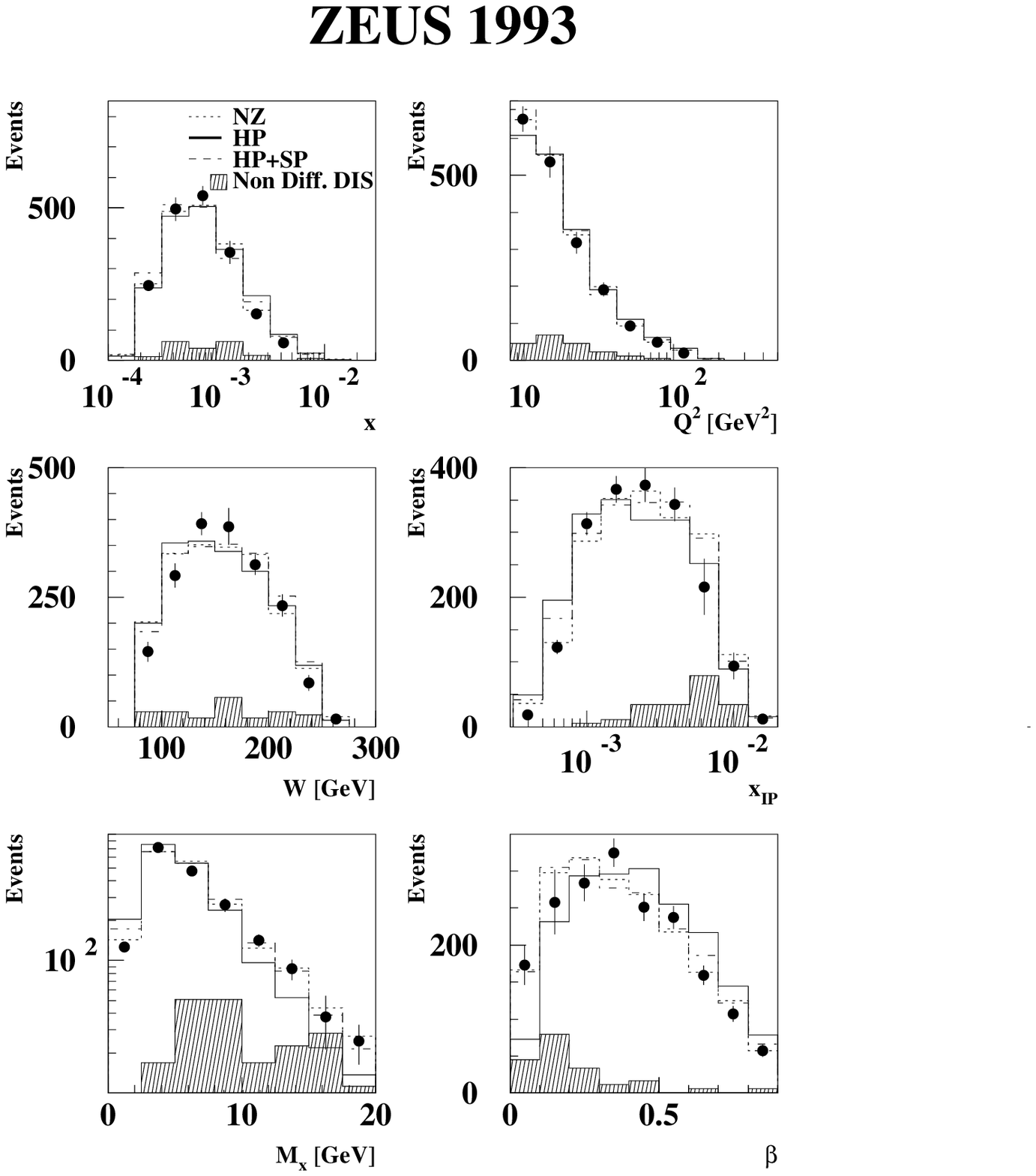}}
\end{center}
\caption{{\protect\small
Observed distributions of $x$, $Q^2$, $W$, $\xpom$, $M_X$ and $\beta$
for the selected ($M_X,y$) intervals.
Uncorrected data are indicated by the dots.
The errors are the statistical errors combined in quadrature
with 50\% of the non-diffractive DIS background.
The predictions from HP (full line), HP+SP (dashed line) and NZ (dotted line)
models are shown.
The non-diffractive DIS background which has been subtracted from the
data is indicated by the shaded area.}}
\label{all}
\end{figure}

Figure~\protect\ref{all} shows the $x$, $Q^2$, W, $\xpom$, $M_X$ and
$\beta$ distributions after applying the \etamax-\thetah~cut, requiring
the data to be in the accepted ranges of ($M_X$, $y$) and subtracting the
DIS background indicated in the figure. The errors on the data points are
calculated by summing in quadrature the statistical error (which is the
dominant error) and 50\% of the total subtracted DIS background (which is
taken as a conservative estimate of the uncertainty due to the DIS
background). In addition, the predictions from the two diffractive models
(NZ and HP) are shown.

In general, both models describe the data. Differences are observed in the
$M_X$ and $\beta$ distributions, where the HP model underestimates the
observed number of events at low $\beta$ values and does not reproduce the
observed $M_X$ distribution at large $M_X$. The NZ model, incorporating a
soft component, describes the observed $\beta$ and $M_X$ distributions.

A pure ``hard" \betan distribution cannot account for the data, therefore,
the observed $\beta$ spectrum was fitted as a sum of a ``hard + soft"
contribution from the POMPYT Monte Carlo. This resulted in a contribution
of $\simeq 60\%$ and $\simeq 40\%$ from HP and SP, respectively. This
HP+SP model is also shown in Fig.~\ref{all}. Comparison with the data
indicates that such a model also describes the observed $\beta$ behaviour.

\begin{figure}[htb]
\begin{center}
\leavevmode
\hbox{%
\epsfxsize = 5.5in
\epsffile{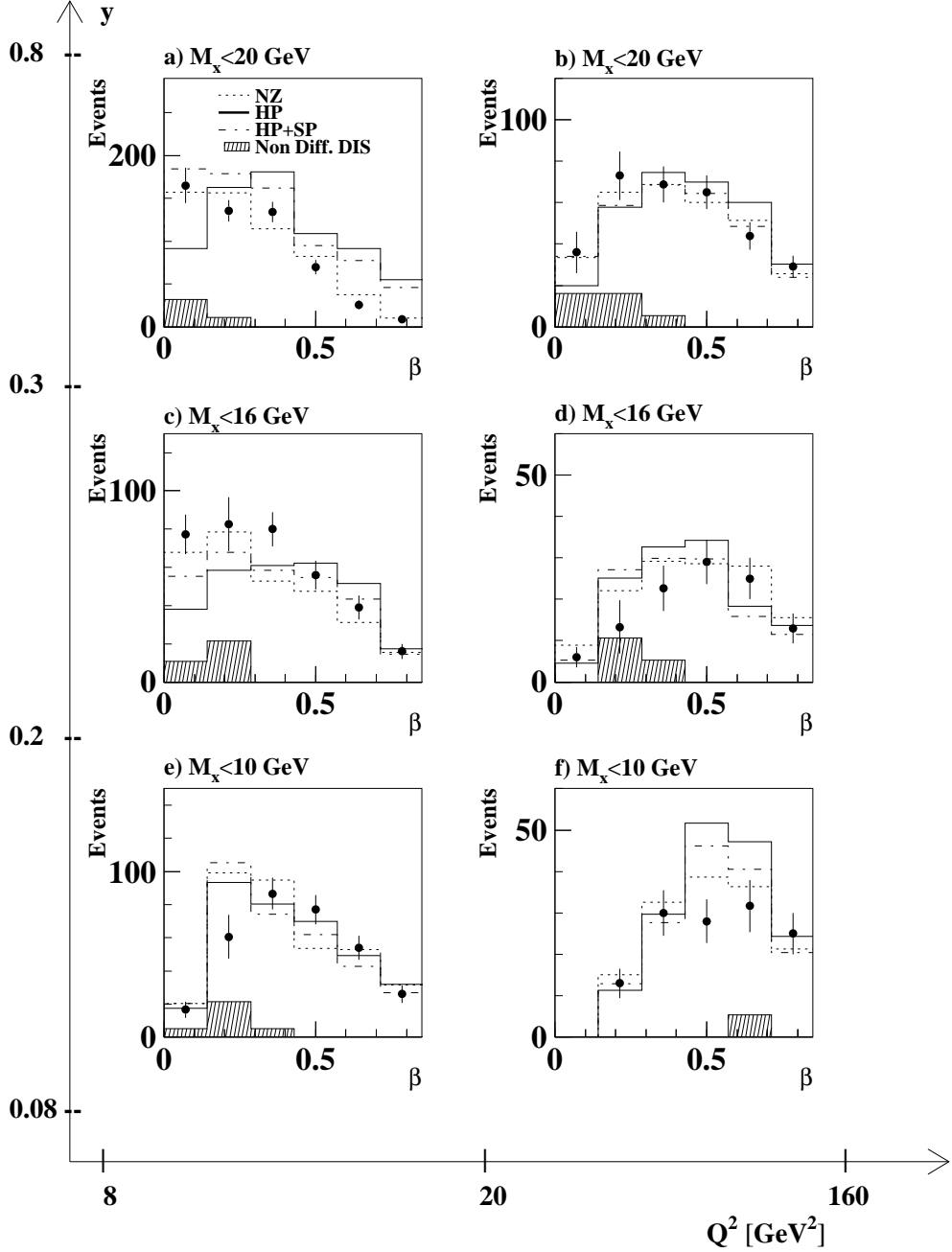}}
\end{center}
\caption{{\protect\small
Observed $\beta$ distribution as a function of ($y,Q^2,M_X$).
The $Q^2$ intervals are 8-20 and 20-160 GeV$^{2}$,
the $y$ intervals are 0.08-0.2, 0.2-0.3 and 0.3-0.8, and
the $M_X$ intervals are (a,b) 0-20, (c,d) 0-16 and (e,f) 0-10 GeV.
Uncorrected data are indicated by the dots.
The errors are the statistical errors combined in quadrature
with 50\% of the non-diffractive DIS background.
The predictions from HP (full line), HP+SP (dashed line)
and NZ (dotted line) models are shown.
The non-diffractive DIS background which has been subtracted from the
data is indicated by the shaded area.}}
\label{dndb}
\end{figure}

To investigate the \betan distribution in more detail,
each ($M_X, y$) interval was divided into two $Q^{2}$ bins:
$$
Q^{2} = 8-20,\ 20-160~{\rm GeV}^2
$$
The results, together with the predictions from the diffractive
models are shown in Fig.~\ref{dndb}.
In general, the Monte Carlo models reproduce the shape of the data
reasonably well.
However, in the high-$y$ and low-$Q^2$ intervals where the mass extends to
larger values (Fig.~\ref{dndb}a and c), the soft contribution is important.
The NZ model describes the data best in this region. The HP+SP model
reasonably describes the data and gives an improved description compared
to the HP model in each ($M_X, y$) interval.

\begin{figure}[htb]
\begin{center}
\leavevmode
\hbox{%
\epsfxsize = 6.5in
\epsffile{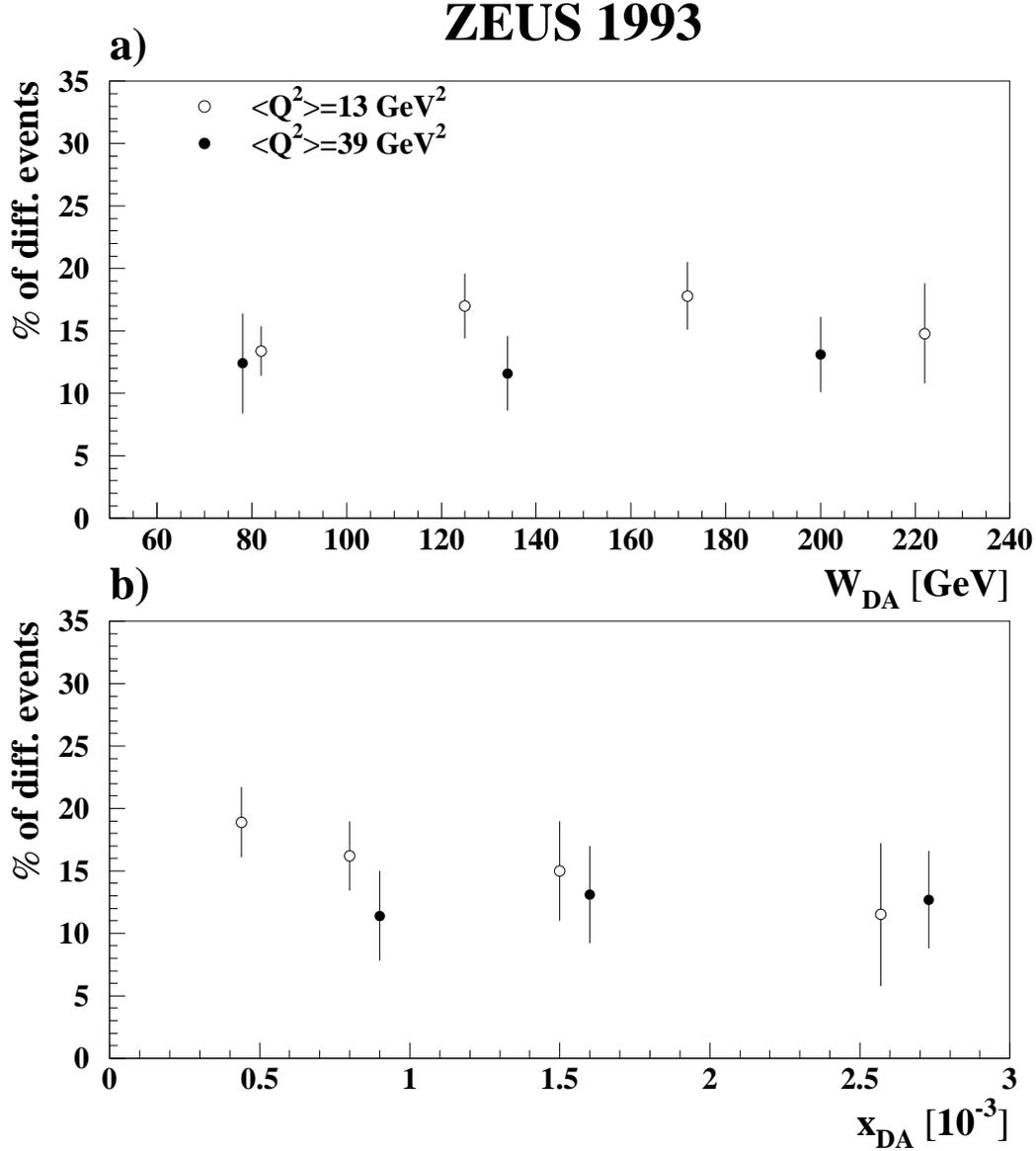}}
\end{center}
\caption{{\protect\small
Observed fraction of diffractive events as a function of $W_{DA}$ and
$x_{DA}$ in two $Q^2$ intervals.
The data are fitted to the NZ model for diffractive processes
and the CDMBGF model for the non-diffractive contribution.
The errors are the statistical errors combined in quadrature
with 50\% of the non-diffractive DIS background.
%In figures c) and d) the data are fitted to the HP model.
}}
\label{figfive}
\end{figure}

Using the NZ model, the combined fit to the \etamax\ and \thetah\
distributions was performed in bins of $W$ and $x$ respectively,
separately for the two $Q^2$ intervals indicated above, to extract the
fraction of diffractive events as a function of these variables.
Figure~\protect\ref{figfive}\ shows the diffractive fraction as a function
of $W$ and $x$ for different values of $Q^2$.  The results extracted using
the HP+SP model agree within statistical errors.  The results extracted
using the HP model give a normalisation which is $\simeq$ 30\% lower, but
with the same dependence on $x$, $W$ and $Q^2$. The fits are mainly
sensitive to the hard component: a large uncertainty on the diffractive
contribution to the DIS sample comes from the soft part in the pomeron
structure function, which is suppressed by the applied cuts, especially at
small values of $W$.  In all cases, no strong dependence of this ratio is
observed as a function of $x$, $W$ or $Q^2$.

\section{Measurement of the diffractive structure function}

As described in section~\ref{s:dsf}, the differential cross section can be
expressed in terms of the diffractive structure function \F2Diff as a
function of $\beta$, $\xpom$ and $Q^2$.  In this section, we discuss the
resolution of the measured quantities and describe the kinematic region
chosen.  We finally discuss the systematic errors and present the results
of the measurement of \F2Diff.

\subsection{Extraction of \F2Diff}
\label{s:f2d}

According to Monte Carlo studies (see section~\ref{s:MC}), the resolution
of $Q^2$ is 25\%, independent of $Q^2$. The resolution of $x$ varies
smoothly with $x$ from 20\% at $x = 10^{-2}$ to 50\% at $10^{-3}$, almost
independent of $Q^2$. The resolution of $M_X$, reconstructed with the
method described in section~\ref{s:kin}, is approximately 27\%,
independent of $M_X$. The $M_X$ reconstruction is affected by energy loss
in inactive material in front of the calorimeter and the position
determination of hadrons. In order to reduce migrations at small masses,
the cell energy thresholds for isolated cells were increased. Monte Carlo
studies show that, except for very small masses ($<$ 3~GeV) where
calorimeter noise becomes important, $M_X$ is systematically shifted by
10\% to smaller values, independent of $y$ and $Q^2$. In order to
compensate for this shift, a correction factor of 1.10 was applied to the
measured $M_X$ values for the determination of the diffractive structure
function.  The resolution of $\xpom$ is approximately 25\%. The resolution
of \betan varies smoothly with \betan from 40\% at $\beta=0.1$ to 20\% at
$\beta=0.8$.

Below $Q^2$ of 8~GeV$^2$, the event acceptance drops below 50\% due to the
box cut requirement. The statistics of the 1993 data allow four ranges in
$Q^2$ to be selected above this lower limit. The migration of events is
large at small values of $x$: we therefore chose bins where the central
$x$-value is above $4 \cdot 10^{-4}$.  The acceptance of the diffractive
component increases as a function of $y$: we therefore select only bins
with $y > 0.08$.  The overall acceptance due to the DIS and diffractive
cuts in the selected bins given in Table~2 is always greater than 50\% and
typically $\simeq$~80\%. The $M_X$ resolution determines the chosen bin
size in the variables $\beta$ and $\xpom$. The purity, defined as the
fraction of simulated events generated in a bin and measured in the same
bin, is always greater than 25\% and typically $\simeq$~40\% in each of
the selected bins.

In order to control the influence of photoproduction background, radiative
corrections and $F_L$ contributions, we restrict our analysis to $y <
0.5$.  As a consequence the minimum scattered electron energy requirement
is raised to 10~GeV. We checked that our sensitivity to $F_L$ is smaller
than the quoted errors in all bins. Furthermore, the region $\beta < 0.8$
is selected to exclude the region of low masses where vector meson
production is dominant.

The level of photoproduction background is estimated in bins of $\xpom$
and $Q^2$ by fits to the $\delta$ distributions (see \cite{F293} for
details).  Since it is typically $\simeq$ 1\% and always below 4\% we do
not correct for this background.

We select bins with $\xpom < 0.01$ and $\beta > 0.1$ where the
non-diffractive component can be safely estimated.  In each of the bins
the number of events is then evaluated by subtraction of the estimated
number of DIS background events, based on the ARIADNE Monte Carlo program
with the normalisation described in section~6. The contribution of the DIS
background is given in Table~2.

To unfold the effects of acceptance and event migration we used the NZ
Monte Carlo event sample, which gives a good description of our data. For
this initial study we used a one-step matrix unfolding procedure and
applied a bin-centring correction for the quoted \F2Diff values.

\subsection{Systematic errors}
\label{s:systematics}

Several systematic checks were performed to  estimate the uncertainties due to
the selection cuts, background estimate and the unfolding. Systematic errors
due to the DIS event selection  were evaluated in the following way (see
\cite{F293} for a detailed discussion):
\begin{itemize}
\item different algorithms were used to identify the scattered electron
which differ in purity and efficiency. The changes to \F2Diff\ were
below 10\%;

\item
the cut on $E_e^\prime$ was decreased from 10 to 5~GeV to study the
effect of a possible mismatch of the shower profiles of data and Monte Carlo
at small energies. The change of \F2Diff\ was less than 5\% in each bin;

\item the box-cut was changed by 2~cm from the nominal values, to study the
effects of electron position reconstruction at  small angles. This resulted
in changes which were always less than 15\%;

\item
the $\delta$-cut was raised from 35 GeV to 40 GeV,
to study the effect of radiative corrections,
which were not included in the simulations. This resulted
in a general shift of $\simeq 10\%$ towards smaller \F2Diff\ values;

\item the $y_{JB}$-cut was changed from 0.04 to 0.02 and to 0.06. This
affected the region of large $\xpom$ where \F2Diff\ changes by about 10\%.

\end{itemize}

Systematic errors due to the diffractive event selection were evaluated in
the following way: \begin{itemize} \item the effect of a possible mismatch
between the hadronic energy scale in the Monte Carlo and the data was
investigated by shifting the hadron energy scale by 7\% in the Monte Carlo
simulation.  The use of the $DA$ variables resulted in changes on \F2Diff\
which were always smaller than 2\%;

\item the fraction of low-mass events was reduced by 50\%.  Due to migrations
from $\beta > 0.8$, this change influences the small $Q^2$, high $\beta$ bin,
where the values were shifted upwards by $\simeq 10\%$;

\item the HP model was used instead of the NZ simulation for unfolding the
data. Some effect was seen in the small $\beta$-region, where the pomeron
structure functions differ. The changes to \F2Diff\ were typically
$\simeq 10\%$;

\item as a systematic check for the estimate of the DIS background the
$\eta_{\rm max}$-cut was reduced from 2.5 to 2.0 resulting in changes of up to
20\% in the highest $\xpom$ bins. The $\eta_{\rm max}$-cut was also increased
from 2.5 to 3.0 resulting in changes of up to 10\%;

\item
similarly, the \thetah~cut was removed, yielding changes below 5\%;

\item
the cells with $\eta>2.5$ were removed to check the dependence on the
double-dissociative contribution, resulting in changes which were up to 5\%.

\end{itemize}

Overall most of these checks yielded results which agree with the standard
method within statistical errors.  The differences of the DIS and
diffractive systematic checks compared to the standard method were
combined in quadrature to yield the quoted systematic errors.

\subsection{Results}

Table~2 summarises the results for \F2Diff, for the 0.54~pb$^{-1}$ ($\pm
3.5\%$) integrated luminosity. The statistical errors include statistical
uncertainties from the Monte Carlo models used for the unfolding.
\begin{table}[h]
\begin{center}
\begin{tabular}[t]{|c|c|c|c|c||r@{~$\pm$}r@{~$\pm$}r|}\hline
  $Q^2$ &~$\beta$~& \xpom & \#events & \#non-diff.
 & \F2Diff &stat. &sys. \\
  (GeV$^2$) & & & & background & \multicolumn{3}{c|}{ } \\ \hline \hline
 10 & 0.175 & 0.0032 & 54 & 7.1 &9.7 & 1.6 &2.8 \\
 10 & 0.175 & 0.0050 & 32 & 5.2 &5.0 & 1.1 &2.3 \\  \hline
 10 & 0.375 & 0.0013 & 62 & 0.9 &37.7 & 5.2 &6.5 \\
 10 & 0.375 & 0.0020 & 43 & 2.8 &22.0 & 3.7 & 3.7\\
 10 & 0.375 & 0.0032 & 15 & 2.8 &9.2 & 3.0 & 4.5 \\ \hline
 10 & 0.65 & 0.00079 & 56 & 0.9 &47.7 & 8.7 &29.9 \\
 10 & 0.65 & 0.0013 & 20 & 0.9 &29.1 & 7.0 & 8.5 \\
 10 & 0.65 & 0.0020 & 23 & 0 &10.9 & 2.3 & 6.9 \\ \hline \hline
 16 & 0.175 & 0.0032 & 48 & 5.2 &9.5 & 1.6 &2.1 \\
 16 & 0.175 & 0.0050 & 50 & 4.7 &6.5 & 1.1 &1.8 \\
 16 & 0.175 & 0.0079 & 33 & 7.5 &3.8 & 0.9 &2.0 \\ \hline
 16 & 0.375 & 0.0013 & 54 & 2.8 &38.2 & 5.9 &5.3 \\
 16 & 0.375 & 0.0020 & 54 & 3.3 &20.1 & 3.1 & 3.6 \\
 16 & 0.375 & 0.0032 & 52 & 3.3 &13.3 & 2.0 &3.6 \\
 16 & 0.375 & 0.0050 & 44 & 3.8 &6.2 & 1.0 &1.8 \\ \hline
 16 & 0.65 & 0.00079 & 49 & 0   &39.8 & 11.6 &13.8 \\
 16 & 0.65 & 0.0013 & 38 & 2.8 &32.5 & 6.3 &6.5 \\
 16 & 0.65 & 0.0020 & 43 & 1.4 &13.3 & 2.5 & 3.7 \\
 16 & 0.65 & 0.0032 & 29 & 0 &8.5 & 1.6 &2.3 \\ \hline \hline
 28 & 0.175 & 0.0050 & 35 & 3.3 &6.4 & 1.3 &1.4 \\
 28 & 0.175 & 0.0079 & 32 & 8.0 &3.8 & 0.9 &1.7 \\ \hline
 28 & 0.375 & 0.0020 & 26 & 1.4 &23.4 & 5.0 & 3.3 \\
 28 & 0.375 & 0.0032 & 35 & 1.9 &15.7 & 2.9 & 2.0 \\
 28 & 0.375 & 0.0050 & 41 & 3.3 &7.5 & 1.3 &1.5 \\
 28 & 0.375 & 0.0079 & 19 & 3.3 &3.1 & 0.9 &1.1 \\ \hline
 28 & 0.65 & 0.0013 & 30 & 0.5 &26.5 & 6.4 &9.4 \\
 28 & 0.65 & 0.0020 & 35 & 1.9 &15.7 & 3.4 & 2.5 \\
 28 & 0.65 & 0.0032 & 25 & 1.4 &9.2 & 2.1 &2.5 \\
 28 & 0.65 & 0.0050 & 23 & 1.4 &5.4 & 1.3 &2.9 \\ \hline \hline
 63 & 0.375 & 0.0050 & 17 & 2.4 &6.8 & 2.0 &1.7 \\
 63 & 0.375 & 0.0079 & 16 & 3.8 &2.6 & 0.9 &1.5 \\ \hline
 63 & 0.65 & 0.0032 & 22 & 0.5 &10.8 & 2.9 &0.8 \\
 63 & 0.65 & 0.0050 & 17 & 0.5 &6.2 & 1.7 &0.9 \\
 63 & 0.65 & 0.0079 & 11 & 2.4 &3.0 & 1.2 &0.7 \\ \hline
\end{tabular}
\end{center}
\label{f2d}
\caption{ZEUS 1993 \F2Diff\ results.
The overall normalisation uncertainty of 3.5\% is not included.
The data contain an estimated 15$\pm$10\% fraction of double-dissociative
events.
}
\end{table}

\begin{figure}[htb]
\begin{center}
\leavevmode
\hbox{%
\epsfxsize = 6in
\epsffile{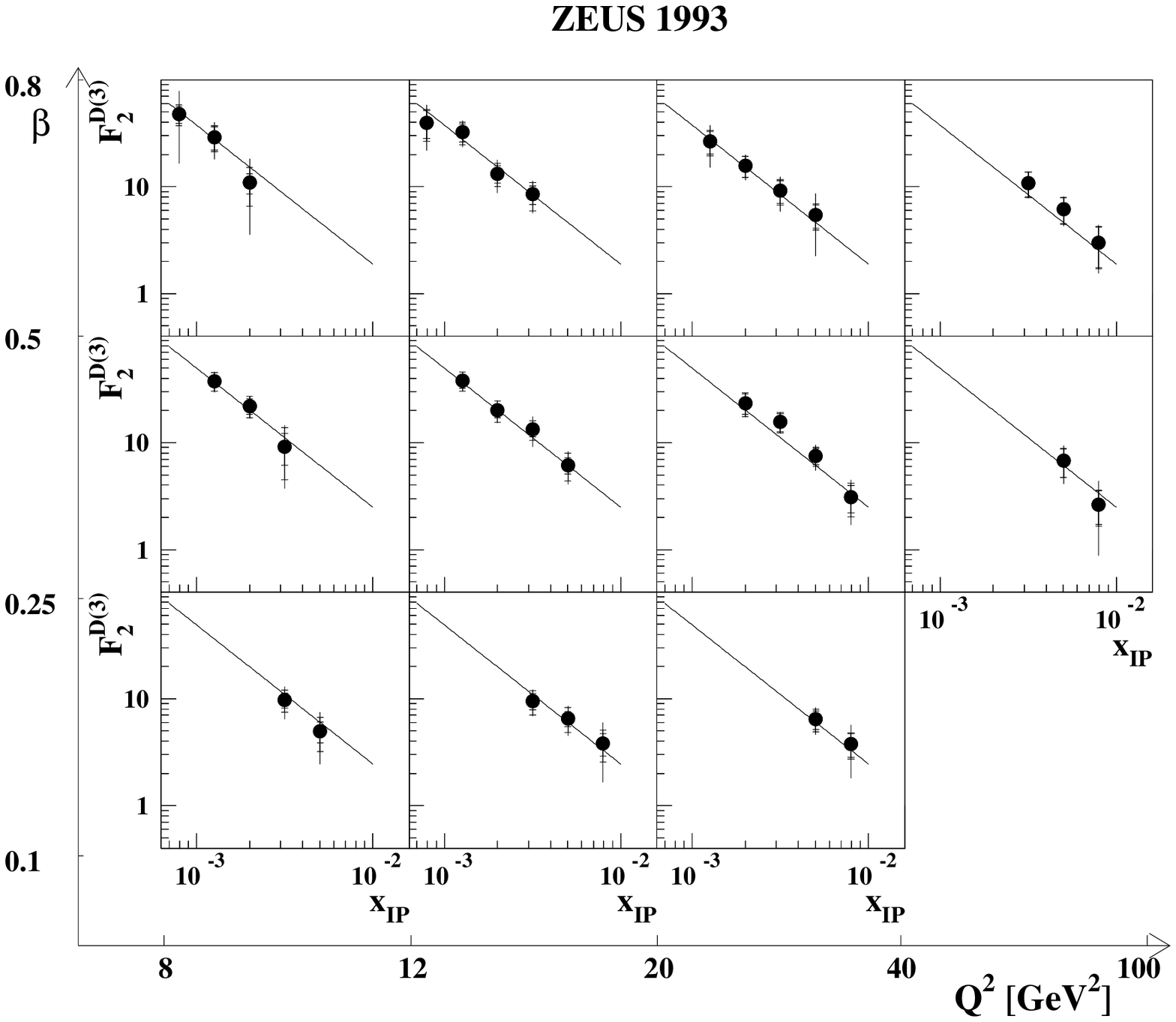}}
\end{center}
\caption{{\protect\small
The results of \F2Diff($\beta, Q^2, \xpom$)
compared to the parametrisation discussed in the text.
The inner error bars show the statistical errors,
the outer bars correspond to the statistical and
DIS event selection systematic errors added in quadrature,
and the full line corresponds to the statistical and
total systematic errors added in quadrature.
Note that the data include an estimated 15\% contribution
due to double dissociation.
The overall normalisation
uncertainty of 3.5\% due to the luminosity uncertainty is not included.
}}
\label{fit}
\end{figure}

The \F2Diff\ results are displayed in Fig.~\ref{fit}.
The data are observed to fall rapidly as a function of increasing $\xpom$.
In the measured bins the dependence of \F2Diff\ on $Q^2$ at fixed
$\beta$ values is weak.
We have investigated whether the $\xpom$-dependence of \F2Diff\ is the same in
each $\beta,Q^2$ interval,
as expected if factorisation holds. For this purpose we performed
fits of the form:
$$ b_i \cdot \xpoma $$
where the normalisation constants $b_i$ were allowed to differ,
while the exponent was the same for each $\beta,Q^2$ interval.
The result of the fit was:
$$ a = 1.30 \pm 0.08~(stat)~^{+~0.08}_{-~0.14}~(sys). $$
The systematic errors are calculated by re-fitting the
\F2Diff\ values according to the variations listed in
section~\ref{s:systematics} and combining the positive or negative
deviations from the central value of $a$ in quadrature.
The overall statistical $\chi^2$ values of these fits are in the range
8.2--14.0 for 23 degrees of freedom depending on the systematic check.
The $\chi^2$ values for each of the $\beta$,$Q^2$ intervals
are in the range 0.1--1.1 per degree of freedom.
Within the present accuracy, the data are therefore consistent
with the assumption of factorisation in the measured kinematic range.
The value of $a$ is consistent with recent
results from the H1 Collaboration of
\mbox{$a = 1.19 \pm 0.06 \pm 0.07$~\cite{h1new}.}

The observed dependence on $\xpom$ is steeper but still
compatible with
a Donnachie-Landshoff type of flux factor which yields $a \simeq 1.09$
and which is based on a phenomenological description of ``soft hadronic''
diffractive interactions.
The modified flux of Goulianos yields a \xpoma dependence
with $a \simeq 0.93$, a value which is disfavoured by the data.

\begin{figure}[htb]
\begin{center}
\leavevmode
\hbox{%
\epsfxsize = 6in
\epsffile{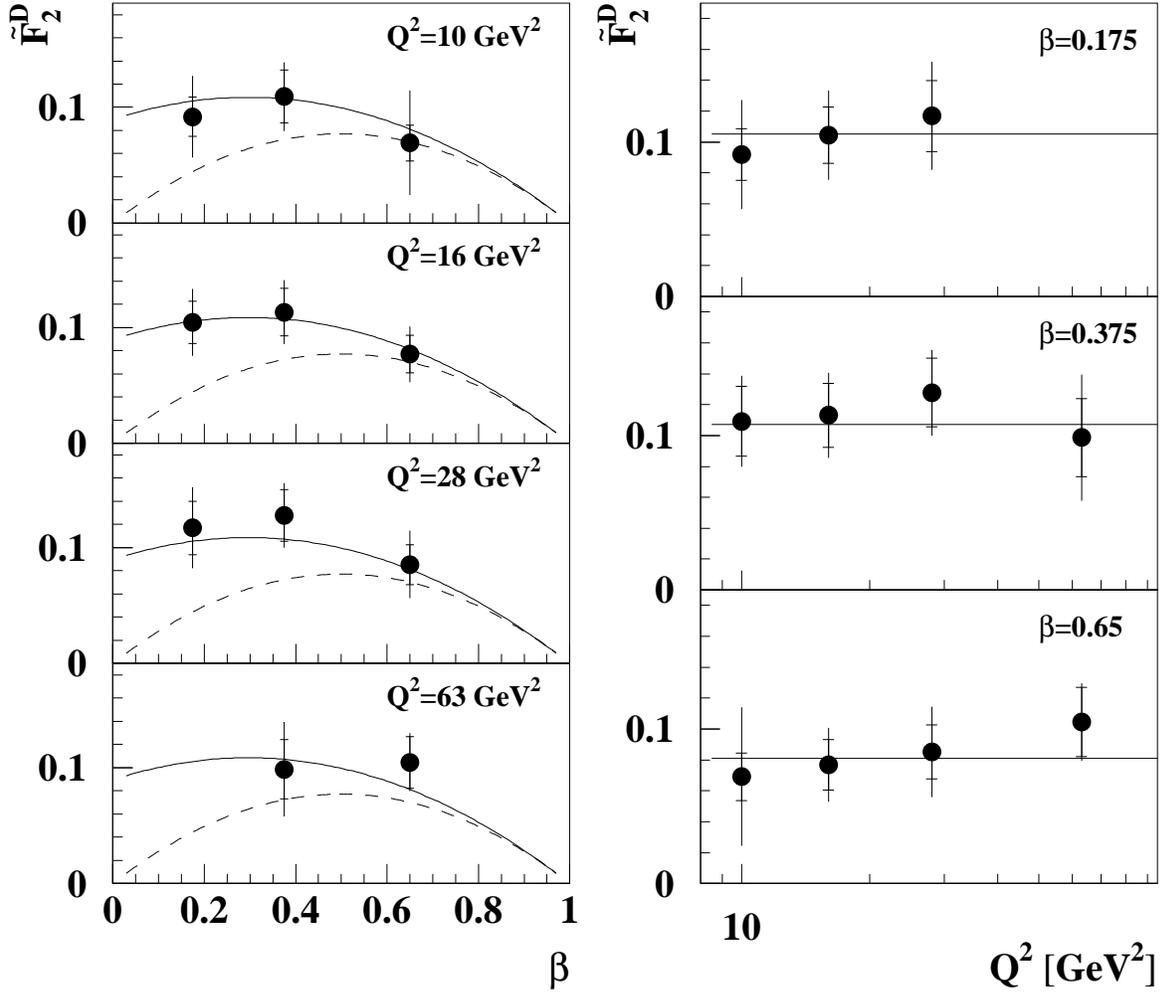}}
\end{center}
\caption{{\protect\small
The results of $\tilde{F}_2^D$($\beta, Q^2$) compared to
the parametrisation discussed in the text, indicated by the full line,
and the $\beta(1-\beta)$ hard contribution, indicated by the dashed line.
The inner error bars show the statistical errors,
the outer bars correspond to the statistical and
systematic errors added in quadrature.
The systematic errors combine in quadrature the fits of the
$\xpom$ dependence due to each of the systematic checks discussed
in the text.
Note that the overall normalisation is arbitrary and is determined by the
experimental integration limits over $\xpom$
($6.3\cdot 10^{-4} <$ \xpom\ $<10^{-2}$).
The data include an estimated 15\% contribution due to double dissociation.
}}
\label{f2pom}
\end{figure}

In order to illustrate the $\beta$ and $Q^2$ dependence of
\F2Diff($\beta, Q^2, \xpom$),
we integrated \F2Diff over the measured range of
\xpom, $6.3\cdot 10^{-4} <$ \xpom\ $<10^{-2}$, using the fitted
\xpom\ dependence.
The resulting values of $\tilde{F}_2^D$($\beta, Q^2$) are shown in
Fig.~\ref{f2pom} as a function of $\beta$ and $Q^2$.
It should be noted that these results assume that a universal \xpom\
dependence holds in $all$ regions of $\beta$ and $Q^2$. In particular,
there is a contribution due to regions of \xpom\ which are not measured
and where the hypothesis of a universal \xpom\ dependence has not been
tested experimentally.

The $\tilde{F}_2^D$($\beta, Q^2$) values as a function of
$\beta$ for fixed $Q^2$ are
consistent with a flat $\beta$ dependence as expected
from the aligned jet model~\cite{bjorken} and the model of
Buchm\"uller~\cite{buchmuller}.
As a function of $Q^2$ for fixed $\beta$, the $\tilde{F}_2^D$($\beta, Q^2$)
values are approximately independent of $Q^2$ for all $\beta$ values,
which is consistent with a picture where the
underlying interaction is the scattering of a virtual photon with
point-like quarks within the pomeron.

As a next step we determined a compact parametrisation for the
\F2Diff\ results, which is also shown in Fig.~\ref{fit} and Fig.~\ref{f2pom},
where the following form was adopted:
$$ \F2Diff = (1/\xpom)^{a} \cdot b \cdot (\beta (1-\beta) + \frac{c}{2}
\cdot (1-\beta)^2), $$
with $a = 1.30$.
This parametrisation assumes factorisation and no $Q^2$-dependence.
The soft contribution to the structure function was considered by the
inclusion of the $(1-\beta)^2$ term.
The multiplicative factor of $\frac{c}{2}$ was chosen such that
the integral over $\beta$ of the soft contribution is equal
to that of the hard contribution when $c = 1$.
The power of 2 was adopted from the NZ model;
this assumption cannot be tested with the current measurement.
The results of the fit were:
$$ b = 0.018 \pm 0.001~(stat) \pm 0.005~(sys), $$
$$ c = 0.57 \pm 0.12~(stat) \pm 0.22~(sys), $$
with a statistical $\chi^2$ in the range 15--23 for 33 degrees of freedom
depending on the systematic check.
A fit without the $(1-\beta)^2$ soft contribution resulted in
$\chi^2$ values in the range 56--81 for 34 degrees of freedom.
This increased $\chi^2$ value indicates that a soft component
is required in the pomeron structure function.

\begin{figure}[htb]
\begin{center}
\leavevmode
\hbox{%
\epsfxsize = 6in
\epsffile{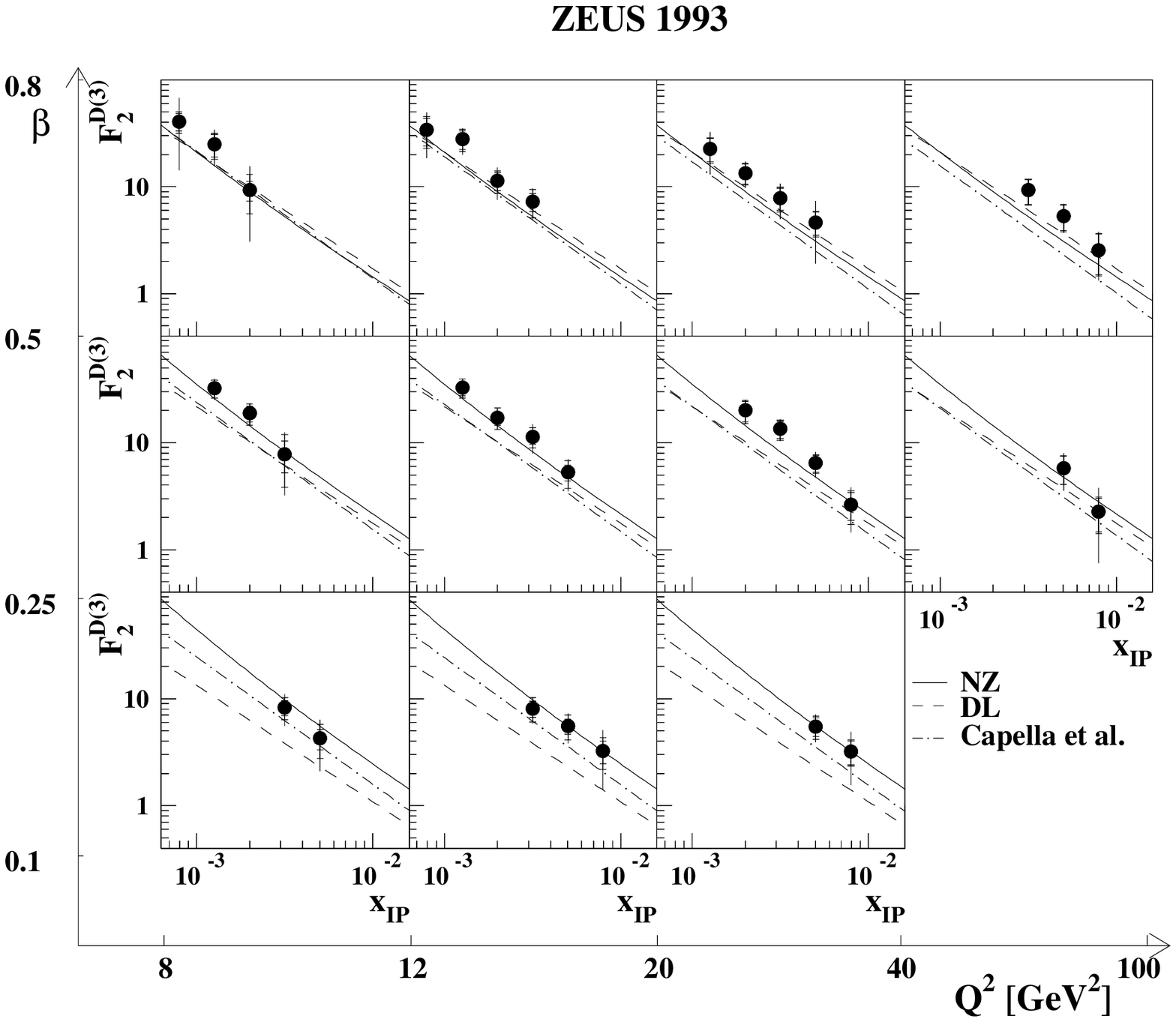}}
\end{center}
\caption{{\protect\small
The results of \F2Diff\ compared to various models discussed in the text.
Note that the estimated 15\% contribution due to double dissociation
has been subtracted in order to compare with models for the single dissociation
cross section.
The inner error bars show the statistical errors,
the outer bars correspond to the statistical and
DIS event selection systematic errors added in quadrature,
and the full line corresponds to the statistical and
total systematic errors added in quadrature.
The overall normalisation
uncertainty of 3.5\% due to the luminosity and 10\% due to the subtraction
of the double dissociation background is not included.
}}
\label{comp}
\end{figure}

We now consider the cross section
predictions of the models discussed in section~\ref{s:MC}.
The \F2Diff\ results are displayed in Fig.~\ref{comp} where
the data are compared
with the predictions of several models of single-diffractive dissociation
for which the momentum sum rule for quarks is not satisfied.
The estimated 15\% fraction of double-dissociative events
has been subtracted in order to compare with these models.

At high $\beta$-values the predictions of Nikolaev-Zakharov,
Donnachie-Landshoff and Capella et al.~underestimate the observed values
slightly, but are generally in reasonable agreement.
At smaller $\beta$-values, the Donnachie-Landshoff parametrisation,
which includes only a hard component of the pomeron structure function,
underestimates the observed \F2Diff. The Capella et al. and
Nikolaev-Zakharov predictions, which also include a soft component,
are able to give a fair description at smaller $\beta$-values.
The factorisation-breaking effects in the model of Nikolaev-Zakharov,
which occur at small $\beta$ values, are too small to be
observable in this analysis.

\begin{figure}[htb]
\begin{center}
\leavevmode
\hbox{%
\epsfxsize = 6in
\epsffile{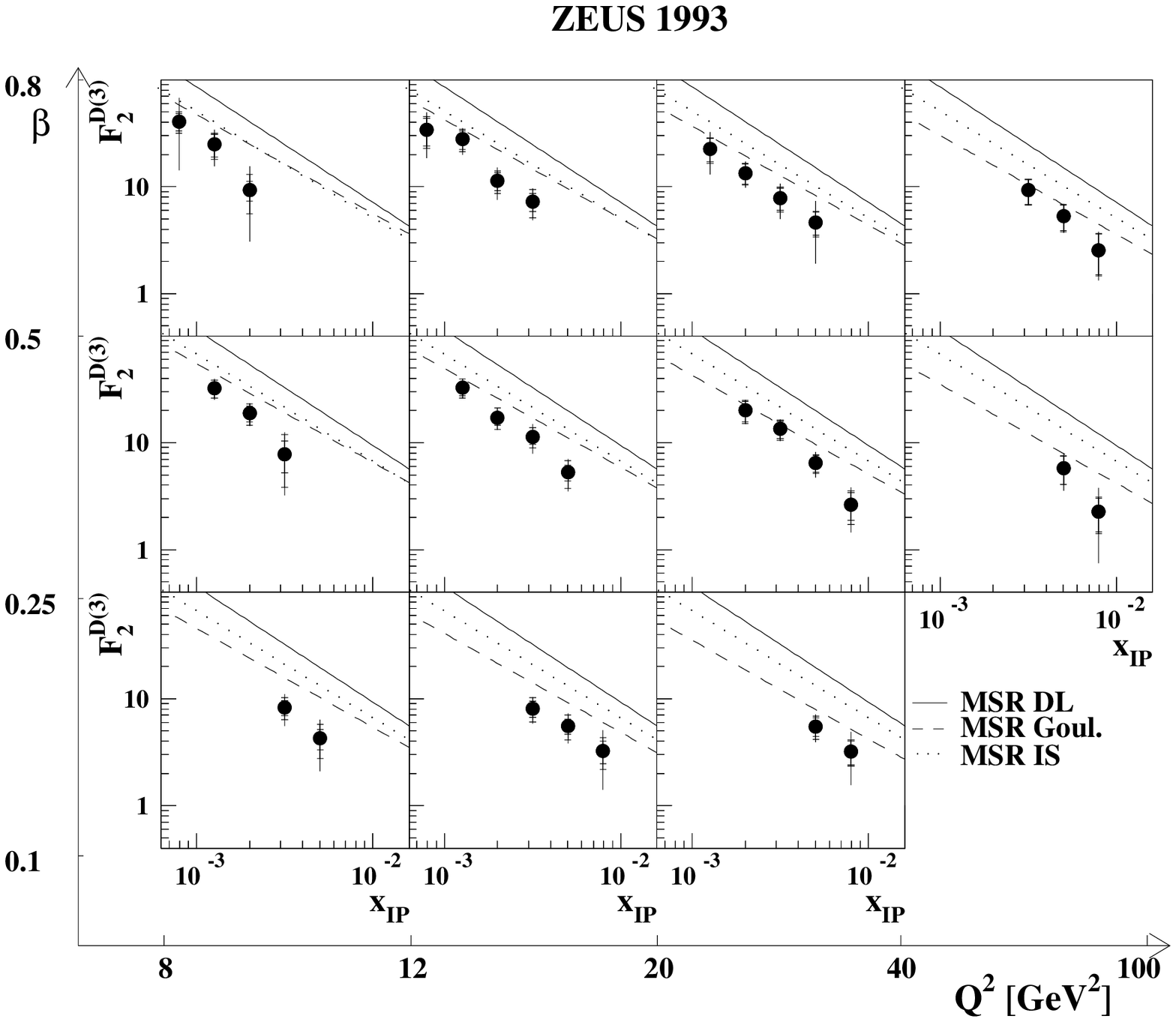}}
\end{center}
\caption{{\protect\small
The results of \F2Diff\ compared to an Ingelman-Schlein type model
for which the momentum
sum rule (MSR) for quarks within the pomeron is assumed.
The $\beta$ dependence is taken from the parametrisation discussed in the text.
Note that the estimated 15\% contribution due to double dissociation
has been subtracted in order to compare with models for the single dissociation
cross section.
The inner error bars show the statistical errors,
the outer bars correspond to the statistical and
DIS event selection systematic errors added in quadrature,
and the full line corresponds to the statistical and
total systematic errors added in quadrature.
The overall normalisation
uncertainty of 3.5\% due to the luminosity and 10\% due to the subtraction
of the double dissociation background is not included.
}}
\label{comp2}
\end{figure}

In Fig.~\ref{comp2} the data are compared with a model for which the momentum
sum rule for the pomeron structure function is assumed for the light
quark flavours (u,d) and the $\beta$ dependence is taken from the
parametrisation, discussed above. Adopting the Donnachie-Landshoff flux
factor, the observed \F2Diff\ is about a factor three to four below the
expectation if the momentum sum rule is assumed to be fulfilled only by
quarks. An uncertainty arises from the choice of the
pomeron flux factor: if the Ingelman-Schlein form for the flux factor is
adopted then the prediction is reduced by approximately 30\%.
Even if the Goulianos prescription for the flux is adopted, the observed
\F2Diff\ results are always below the predictions. These comparisons indicate
that in an Ingelman-Schlein type model the quarks alone inside the pomeron do
not satisfy the momentum sum rule.

\section{Conclusions}

The properties of diffractive DIS events with $Q^2 > 8$ GeV$^2$, selected
by a large rapidity gap requirement, have been investigated.  Different
Monte Carlo models, such as the POMPYT model, with a soft plus a hard
pomeron structure function, or the Nikolaev-Zakharov model describe the
shape of the observed kinematic distributions.  Using these models,
reliable acceptance corrections for the measured data can be obtained and
corrected cross sections can be determined.  The relative contribution of
diffractive events to the total DIS cross section is found to have no
strong dependence on $x$, $W$ or $Q^2$.

The diffractive proton structure function \F2Diff\ is presented,
integrated over $t$, the square of the momentum transfer at the proton
vertex, as a function of $\xpom$, the momentum fraction lost by the
proton, of $\beta$, the momentum fraction of the struck quark with respect
to $\xpom$, and of $Q^2$. The structure function is measured in the
kinematic range of $0.08<y<0.5$, $8<Q^2<100$~GeV$^2$, $6.3\cdot 10^{-4} <$
\xpom\ $<10^{-2}$ and $0.1<\beta<0.8$. Within the experimental errors, the
measurement is consistent with models where diffraction is described by
the exchange of a particle-like pomeron where the structure function
factorises into a pomeron flux factor, which depends on \xpom\ and a
pomeron structure function, which is independent of \xpom. The diffractive
structure function is also well-described by the Nikolaev-Zakharov model,
which does not require the concept of a particle-like pomeron, in terms of
overall normalisation and dependence on the kinematic variables, \xpom,
$\beta$ and $Q^2$. The \xpom\ dependence is consistent with the form
\xpoma\ where $a = 1.30 \pm 0.08~(stat)~^{+~0.08}_{-~0.14}~(sys)$ in all
bins of $\beta$ and $Q^2$. The value of $a$ is slightly higher but
compatible with that obtained from hadron-hadron interactions and in
agreement with recent results from the H1 collaboration. In the measured
$Q^2$ range, the pomeron structure function is approximately independent
of $Q^2$ at fixed $\beta$ consistent with an underlying interaction where
the virtual photon scatters off point-like quarks within the pomeron. The
$\beta$-dependence of the pomeron structure function requires both a hard
and a soft component.  In an Ingelman-Schlein type model, where commonly
used pomeron flux factor normalisations are assumed, it is found that the
quarks within the pomeron do not saturate the momentum sum rule.

\section*{Acknowledgements}

The experiment was made possible by the inventiveness and the diligent
efforts of the HERA machine group who continued to run HERA most
efficiently during 1993.

The design, construction and installation of the ZEUS detector have
been made possible by the ingenuity and dedicated effort of many people
from inside DESY and from the home institutes who are not listed as authors.
Their contributions are acknowledged with great appreciation.

The strong support and encouragement of the DESY Directorate
has been invaluable.

We would like to thank A. Donnachie, L. Frankfurt, G. Ingelman and
N. Nikolaev for valuable discussions.

%--------- REFERENCES -------------

\clearpage
%%%%%%%%%%%%%%%%%%%%%%%%%%%%%%%%%%%%%%%%%%%%%%%%%%%%%%%%%%%%%%%%%%%%

\end{document}